\documentclass[english,aps,prd,showpacs,titlepage,authoryear]{revtex4}
\usepackage{amssymb}
\usepackage{amsmath}
\usepackage{mathtools}
\usepackage{epigraph}
\usepackage{eucal}
\usepackage[dvips]{graphics}
\usepackage{epsfig}
\usepackage{bm}
\usepackage{emerald}
\usepackage{babel}
\usepackage{slantsc}
\usepackage{array}
\usepackage{hyperref}
\usepackage[sort&compress]{natbib}

\def\st{\scriptstyle}
\def\sst{\scriptscriptstyle}
\def\be{\begin{equation}}
\def\ee{\end{equation}}
\def\ba{\begin{eqnarray}}
\def\ea{\end{eqnarray}}

\def\H{{\mathcal H}}

\def\a{\alpha}
\def\b{\beta}
\def\g{\gamma}     \def\G{\Gamma}
\def\d{\delta}

\def\e{\epsilon}

\def\m{\mu}
\def\n{\nu}

\def\t{\tau}

\def\la{\label}
\def\pd{\partial}
\def\le{\left}
\def\ri{\right}
\def\mm{{\mathtt f}}

\begin{document}
\title{Celestial Ephemerides in an Expanding Universe}
\author{Sergei M. Kopeikin}
\affiliation{Department of Physics \& Astronomy, \\
University of Missouri, 322 Physics Bldg., \\Columbia, MO 65211, USA}
\email{kopeikins@missouri.edu}

\begin{abstract}
Post-Newtonian theory of motion of celestial bodies and propagation of light was instrumental in conducting the critical experimental tests of general relativity and in building the astronomical ephemerides of celestial bodies in the solar system with an unparalleled precision. The cornerstone of the theory is the postulate that the solar system is gravitationally isolated from the rest of the universe and the background spacetime is asymptotically flat. The present article extends this theoretical concept and formulates the principles of celestial dynamics of particles and light moving in gravitational field of a localized astronomical system embedded to the expanding Friedmann-Lema\^itre-Robertson-Walker (FLRW) universe. We formulate the precise mathematical concept of the Newtonian limit of Einstein's field equations in the conformally-flat FLRW spacetime and analyze the geodesic motion of massive particles and light in this limit. We prove that by doing conformal spacetime transformations, one can reduce the equations of motion of particles and light to the classical form of the Newtonian theory. However, the time arguments in the equations of motion of particles and light differ from each other in terms being proportional to the Hubble constant $\H$. This leads to the important conclusion that the equations of light propagation used currently by Space Navigation Centers for fitting range and Doppler-tracking observations of celestial bodies are missing some terms of the cosmological origin that are proportional to the Hubble constant $\H$. We also analyze the effect of the cosmological expansion on motion of electrons in atoms. We prove that the Hubble expansion does not affect the atomic frequencies and, hence, does not affect the atomic time scale used in creation of astronomical ephemerides.
We derive the cosmological correction to the light travel time equation and argue that their measurement opens an exciting opportunity to determine the local value of the Hubble constant $\H$ in the solar system independently of cosmological observations.
\end{abstract}
\maketitle
\epigraph{I praise you, Father, Lord of heaven and earth, because you have hidden these things from the wise and learned, and revealed them to little children.}{Luke,\textit{10:21}}

\section{Introduction}\la{intrd}

Post-Newtonian celestial mechanics is an important branch of modern gravitational physics  \citep{1987thyg.book..128D,2006LRR.....9....3W} including gravitational wave astronomy \citep{2011PNAS..108.5938W}. It is also a practical tool of modern applied astronomy used for precise navigation of spacecrafts in deep space and for calculation of high-precise astronomical ephemerides of celestial bodies in the solar system \cite{2003AJ....126.2687S,2011rcms.book.....K}. The mathematical formalism of the post-Newtonian approximations is getting progressively complicated as one goes from the Newtonian limit to higher orders \citep{2006LRR.....9....4B,2011mmgr.book..167S}. For this reason the theory has been developed under a basic assumption that the background spacetime is asymptotically flat. Mathematically, it means that the full spacetime metric, $g_{\a\b}$ (the Greek sub-indices $\a,\b,\g,\ldots$ takes values 0,1,2,3), is decomposed around the background Minkowskian metric, $\mm_{\a\b}={\rm diag}(-1,1,1,1)$, into a linear combination
\be\la{i1}
g_{\a\b}=\mm_{\a\b}+h_{\a\b}\;,
\ee
where the perturbation, $h_{\a\b}$, is the post-Newtonian series with respect to the powers of $1/c$, where $c$ is the ultimate speed in nature (the speed of light in vacuum). Post-Newtonian approximations is the method to determine $h_{\a\b}$ by solving Einstein's field equations with the tensor of energy-momentum  of matter of a localized astronomical system $\mathfrak{T}_{\a\b}$ as a source of the field, by iterations starting from $h_{\a\b}=0$ in the expression for $\mathfrak{T}_{\a\b}$. The solution of the field equations and the equations of motion of the astronomical bodies are derived in some inertial coordinates $r^\a=\{ct,{\bm r}\}$ where $t$ is the coordinate time, and ${\bm r}=\{r^i\}$ are spatial coordinates. The post-Newtonian theory in asymptotically-flat spacetime has a well-defined Newtonian limit determined by:
\begin{enumerate}
\item[(1)] equation for the Newtonian potential, $U=c^2 h_{00}/2$,
\be\la{z1}
U(t,{\bm r})=\int_{\cal V}\frac{\rho(t,{\bm r}')d^3r'}{|{\bm r}-{\bm r}'|}\;,
\ee
where $\rho=\mathfrak{T}_{00}$, is the density of matter  producing the gravitational field,
\item[(2)] equation of motion for a massive particle
\be\la{z2}
{\bm r}''={\bm\nabla}U\;,
\ee
where ${\bm\nabla}=\{\pd/\pd x^i\}$ is the operator of gradient, ${\bm r}={\bm r}(t)$ is a radius-vector from the origin of the coordinates to the current position of the particle, and a prime denotes a total derivative with respect to time $t$,
\item[(3)] equation of motion for light (massless particle)
\be\la{z3}
{\bm r}''=0\;,
\ee
which tells us that light moves along a straight line with constant velocity, ${\bm r}={\bm\xi}+c{\bm k}t$, where ${\bm\xi}$ is the impact parameter of the light ray with respect to the origin of the coordinates, and ${\bm k}$ is a unit vector directed along the direction of propagation of the light ray \citep[section 7.3]{2011rcms.book.....K}.
\end{enumerate}
These equations are considered as fundamentals for creation of astronomical ephemerides of celestial bodies in the solar system \citep{2011rcms.book.....K} and in any other localized system of self-gravitating bodies like a binary pulsar \citep{2005hpa..book.....L}. In all practical cases they have to be extended to take into account the post-Newtonian corrections sometimes up to the 3-d post-Newtonian order of magnitude \citep{2011PNAS..108.5938W}. It is important to notice that in the Newtonian limit the coordinate time $t$ of the gravitational equations of motion (\ref{z2}), (\ref{z3}) coincides with the proper time of observer $\t$ that is practically measured with an atomic clock. In other words, the Newtonian limit of general relativity taken in the asymptotically-flat background spacetime equates the three time scales entering the Newton's law of gravity (\ref{z2}), the equation of light propagation (\ref{z3}), and the equation of motion of an electron in atoms which solution defines the frequency of atomic transitions between different energy levels inside the atom that are used in practical realization of SI second \citep{2004CRPhy...5..799K}.

The post-Newtonian theory is well-established, mathematically elegant and self-consistent. It passed through numerous experimental tests with a flying color \citep{2006LRR.....9....3W}. Is there any problem which we should worry about? The answer is ``yes'' and the problem is that the background is not asymptotically-flat Minkowskian spacetime but is described by the FLRW metric, $\bar g_{\a\b}$. We live in the expanding universe. Therefore, the right thing would be to replace the post-Newtonian decomposition (\ref{i1}) with the more adequate post-Friedmanian one
\be\la{i2}
g_{\a\b}=\bar g_{\a\b}+\varkappa_{\a\b}\;,
\ee
where $\varkappa_{\a\b}$ is the metric perturbation around the cosmological background being represented as a series not only with respect to the  post-Newtonian parameter $1/c$ but also, with respect to the Hubble parameter, $\H$. The latter terms appear in the post-Friedmanian series as a result of the Hubble expansion of the universe. Generalization of the post-Newtonian theory on the Minkowski background spacetime to that on the expanding background manifold is tantalizing and was a matter of considerable efforts of many researchers \citep{1945RvMP...17..120E,1946RvMP...18..148E,1954ZPhy..137..595S,2000CQGra..17.2739B,1933MNRAS..93..325M,2004CeMDA..90..267K,2007CQGra..24.5031M}. A recent article \citep{2010RvMP...82..169C} summarizes the previous results and provides the reader with a number of other valuable resources.

We notice that most of the previous works on celestial mechanics in cosmology used joining of two (for example, Schwarzschild and Friedmann) spherically-symmetric exact solutions of Einstein's equations, that was achieved in many different ways. McVittie's solution \citep{1933MNRAS..93..325M} is perhaps the most successful mathematically but yet lacks a clear physical interpretation \citep{2010RvMP...82..169C}. What is still missing in literature is the precise mathematical formulation of the Newtonian limit for a self-gravitating localized astronomical system (a binary star, a solar system, a galaxy) having no specific spacetime symmetry while embedded to the expanding universe and coupled through the gravitational interaction with the time-dependent background geometry. Theoretical description of the Newtonian limit for a localized astronomical system in expanding universe should conform with the results of the post-Newtonian theory obtained in the asymptotically-flat spacetime. Such a description will allow us to directly compare the equations of the ``canonical'' post-Newtonian celestial mechanics to its cosmological counterpart. Therefore, the task is to derive a set of the Newtonian-like equations in cosmology in some coordinates introduced on the background manifold, and to map them onto the set of the Newtonian equations (\ref{z1})--(\ref{z3}) in asymptotically-flat spacetime. In addition, the equations of motion of electron in atom should be also re-formulated to take into account the Hubble expansion in order to check whether it appears in the equations and affects the accuracy and stability of the atomic time or not. All together such a theory of the post-Newtonian celestial mechanics and electrodynamics would be of a paramount importance for extending the tools of experimental gravitational physics to the field of cosmology and astrophysics, for example, to derive the cosmological extension of the PPN formalism \citep{2006LRR.....9....3W}. The present article discusses the main ideas and principal results of such a theoretical approach.

The problem of a paramount importance here is to find out whether the Newtonian limit of the cosmological perturbation approximates the Newtonian theory so that it contains only terms of the quadratic order in $\H$, or some terms being linear with respect to $\H$, survive. The same question should be answered for the orbital motion of electrons in atoms which, at least in principle, might be affected by terms being linearly proportional to the Hubble parameter.
We will prove that the Newtonian theory for massive particles and the Coulomb approximation for electric charges can be indeed approximated up to the terms of the order of $\H^2$ in the same coordinate chart. However, this conclusion turns out to be  inapplicable to the propagation of light. If equations of motion of electromagnetic signal are formulated in the same coordinates as those for massive particles and charges, they show the presence of terms which are linear with respect to $\H$. The reason behind this intriguing difference is rather simple -- Maxwell's equations for freely-propagating electromagnetic waves are conformally-invariant with respect to conformal transformation of the metric tensor while the equations of motion of massive particles and charges are not. Therefore, it is impossible to find out a single space and time coordinate transformation in FLRW manifold that will reduce equations of motion of particles, charges, and electromagnetic waves simultaneously to their classic form that they have in the asymptotically-flat spacetime. We need at least two different time scales in order to do this.

This simple mathematical fact leads to a fascinating theoretical conclusion -- the Hubble expansion of the universe does not permit the existence of a single uniform time scale for physically-different gravitational and electromagnetic-wave processes. Time scale in gravitational equations of massive bodies known as ephemeris time (ET) \citep{2011rcms.book.....K}, coincides with the atomic time scale (AT) governed by the orbital motion of electrons in atoms giving access to SI unit of time \citep{2004CRPhy...5..799K}. Both time scales are identical with the cosmological time $t$ of FLRW background universe. However, neither ET nor AT coincide with the time scale defined by a geodesic motion of a free electromagnetic wave under condition that the spatial coordinates defining positions of massive particles, charges, and photons belong to the same chart on the manifold. We show that the time scale of freely-propagating electromagnetic wave diverges quadratically as time passes on from AT and ET.

The idea of using propagation of electromagnetic signal in space as a clock belongs to Marzke and Wheeler \citep{marweel} who proposed to run a light ray continuously between two mirrors which move along time-like geodesics of a background manifold. The electromagnetic time scale is measured by counting the number of reflections of light from the mirrors. In fact, the motion of mirrors may be not geodesic -- it is sufficient to know their world lines with required precision to eliminate the effect of the Doppler shift on the frequency of light. Marzke-Wheeler's idea has been realized in frequency standards using a microwave cavity or an optical resonator. On the other hand, the same idea is used for a long time in the Doppler-tracking radio technique but it was never associated with Marzke-Wheeler's optical clock. Doppler tracking of frequency of electromagnetic signal whose phase is locked to the primary atomic standard on Earth, allows a coherent measurement of the phase of electromagnetic wave propagating back and forth between a radio emitter on Earth and a spacecraft's transponder. Thus, the Doppler tracking is a high-quality radio resonator having the size of the solar system and formed by two ``mirror'' -- radio emitter on Earth and transponder on board of the spacecraft. The above-mentioned quadratic divergence of the electromagnetic time scale from AT and ET is caused by the Hubble expansion, and the disagreement between the time scales accumulates proportionally to $\H t^2$. If the divergence between the electromagnetic and AT/ET time scales is measured, it will allow to measure the local value of the Hubble constant $\H$ without resorting to cosmological observations.

We argue that the Pioneer effect has, in fact, a simple and natural explanation as the cosmological effect of quadratic divergence between the electromagnetic and atomic time scales governing the propagation of radio wave in the Doppler tracking system and the atomic clock on Earth respectively. It has nothing to do with the ``anomalous acceleration'' of spacecraft which was for a long time a matter of controversial theoretical speculations \citep{2008ASSL..349...75L}. The present paper explains the Pioneer effect in the framework of general relativity without any resort to ``new physics'' or to an alternative theory of gravity. The most precise measurement of the local value of the Hubble constant can be achieved in a near future with a dedicated space mission equipped with precise atomic clocks and Doppler tracking system \citep{2007SPIE.6673E...6L}. The mission must be designed in such a way that minimizes various on-board generated systematic effects like heat, gas leakage, etc. (see section \ref{expHu})

In the following sections we describe the background FLRW manifold and its cosmological perturbation caused by a localized astronomical system. The perturbation is presented as the post-Friedmannian series with respect to the Hubble parameter $\H$. Only the linear terms will be found explicitly. The quadratic in $\H$ terms are important for discussion of how dark matter and dark energy affects the orbital evolution of the localized system but they are not a primary concern of the present article and are given elsewhere \citep{koppetr}. The linearized cosmological perturbation of FLRW manifold is, then, decomposed into post-Newtonian series with respect to the parameter $1/c$.  The Newtonian limit of the perturbation is defined as an ``osculating'' solution which has the Newtonian form in the limit $1/c\rightarrow\infty$ but with the Hubble parameter $\H$ being retained. The gravitational field obtained in this way, is used to derive the equations of motion of particles, and light. Maxwell's equations for electric charge distribution of an atomic nucleus are formulated on FLRW background with taking into account all terms depending of the Hubble parameter. Their solution is found in the form of the retarded integral which is expanded in series with respect to $1/c$ but with the Hubble parameter retained. Equations of motion of an electron orbiting the atomic nucleus are formulated in a covariant form and also decomposed in the series with respect to $1/c$ with the Hubble parameter included. We, finally, discuss the impact of the proposed theory on astronomical ephemerides and argue that it can explain the physical origin of the Pioneer anomaly \citep{2002PhRvD..65h2004A,2010IAUS..261..189A}. The concluding remarks are about the local coordinates on cosmological manifold and the possible missions to measure the Hubble expansion in the solar system.

\section{Notations}\la{notjk}

The notations used in the present paper are as follows:
\begin{itemize}
\item Greek indices run through values $0,1,2,3$, and Latin indices take values $1,2,3$,
\item Einstein's summation rule is used for repeated indices,
\item $t$ and $x^i=\{x,y,z\}$ are the coordinate time and spatial coordinates of the FLRW metric,
\item $x^\a=\{c\eta,x^i\}$ are the conformal coordinates with $\eta$ being a conformal time,
\item bold letters denote spatial vectors in Minkowski spacetime ${\bm x}=\{x^i\}$,
\item a prime $f'=df/dt$ denotes a total derivative with respect to time $t$,
\item a dot $\dot f=df/d\eta$ denotes a total derivative with respect to the conformal time $\eta$,
\item $\pd_\a=\pd/\pd x^\a$ is a partial derivative with respect to a conformal coordinate $x^\a$,
\item a vertical bar, $f_{|\a}$, denotes a covariant derivative of the function $f$ with respect to the background metric,
\item a bar over function, $\bar f$, means that the function belongs to the background spacetime manifold,
\item tensor indices of geometric objects on the background manifold are raised and lowered with the full metric $\bar g_{\a\b}$,
\item tensor indices of geometric objects on the conformal spacetime are raised and lowered with the Minkowski metric $\mm_{\a\b}$,
\item the scale factor of FLRW metric is denoted as $R=R(t)$, or as $a=a(\eta)=R[t(\eta)]$,
\item the Hubble parameter $\H=R'/R$, and the conformal Hubble parameter $H=\dot a/a$,
\item the present paper operates with the WMAP value of the Hubble constant $\H=71.0\pm 2.5$ (km/sec)/Mpc \citep{0067-0049-192-2-14}. It corresponds to $\H=(2.3\pm 0.1)\times 10^{-18}$ s$^{-1}$.
\end{itemize}
The present paper deals with the theory taking into account all terms that are of the first order in the Hubble parameter $\H$. Terms which are proportional to $\H^2$ will be systematically neglected and published somewhere else \citep{koppetr}.

\section{Background manifold}\la{bckman}
\epigraph{The laws of nature evidently obey certain principles of symmetry, whose consequences we can work out and compare with observation.}{Steven Weinberg, {\it Symmetry: A 'Key to Nature's Secrets}}
The background manifold in the present paper is given by FLRW solution of Einstein's equations. The cosmological metric, $\bar g_{\a\b}$, is spherically-symmetric and may admit the three-dimensional curvature of space which is characterized by a constant parameter $k$ that take either of three values, $k=\{-1,0,1\}$. Cosmological observations consistently indicate that $k=0$ \citep{0067-0049-192-2-14}, and we accept this value in the present paper. The background cosmological metric with $k=0$ is
\be\la{i3}
ds^2=-c^2dt^2+R^2(t)\le(dx^2+dy^2+dz^2\ri) \;,
\ee
where, $c=299 792 458$ m$\cdot$s$^{-1}$, is the fundamental speed, $R(t)$ is the dimensionless scale factor depending only on the cosmological time $t$, and $x^i=\{x,y,z\}$ are spatial coordinates. A continuous set of fiducial observers having fixed values of the space coordinates form the Hubble flow. Each Hubble observer measures the proper time $\t$ defined by $d\t^2=-c^2ds^2$. For fixed spatial coordinates, $\t=t$, which allows us to derive time $t$ if the proper time $\t$ is measured with the help of some ideal clock. Usual assumption is that the proper time can be practically accessed with the help of atomic clocks. In expanding universe this assumption should be proved by deriving equation of motion of an electron in atom and demonstrating that the resulting equation can be reduced to the same form as in asymptotically-flat spacetime. We provide such a proof in section \ref{atpr}.

It is also important to realize that the clocks of the Hubble observers are synchronized in the sense of the Einstein synchronization procedure which is based on the exchange of electromagnetic (light) signals \citep{1972gcpa.book.....W,LL}. Hypersurfaces of simultaneity for the Hubble observers are defined by the condition $t={\rm const.}$. The proper distance, $\ell$, on the hypersurface of simultaneity is $\ell=R(t)\sqrt{x^2+y^2+z^2}$.

For calculational purposes it is more convenient to introduce conformal coordinates $x^\a=(c\eta,x,y,z)$. The conformal time $\eta$ is related to the cosmological time $t$ by an ordinary differential equation
\be\la{i4}
\frac{dt}{d\eta}=a(\eta)\;,
\ee
where $a(\eta)\equiv R[t(\eta)]$ is the scale factor of the spatial part of the metric (\ref{i3}) expressed as a function of $\eta$. The background metric $\bar g_{\a\b}$ in the conformal coordinates is conformally-flat. It may be worth mentioning that the conformally-flat and asymptotically-flat spacetimes are not physically identical \citep{2011GReGr..43..901P}.
\be\la{i5}
\bar g_{\a\b}=a^2(\eta)\mm_{\a\b}\;.
\ee
FLRW spacetime manifold with the metric (\ref{i5}) has the Christoffel symbols,
\be\la{i6}
c\bar\G^\a{}_{\b\g}=-\H\le(\d^\a_\b\bar u_\g+\d^\a_\g\bar u_\b-\bar u^\a\bar g_{\b\g}\ri)\;,
\ee
and the Ricci tensor
\be\la{i6a}
c^2\bar R_{\a\b}=\H'\le(\bar g_{\a\b}-2\bar u_\a\bar u_\b\ri)+3\H^2\bar g_{\a\b}\;,
\ee
where $\H=a'/a$, $\bar u^\a=a^{-1}\d^\a_0$ is a four-velocity of the Hubble flow, and $\bar u_\a=\bar g_{\a\b}\bar u^\b=-a\d^0_\a$, and a prime denotes a derivative with respect to the Newtonian time $t$.

The dynamic evolution of the background manifold is determined by Einstein's equations
\be\la{i7}
\bar G_{\a\b}=\frac{8\pi G}{c^4}\bar T_{\a\b}\;,
\ee
where
\be\la{i7a}
\bar G_{\a\b}=\bar R_{\a\b}-1/2\bar g_{\a\b}\bar R^\m{}_{\m}\;,
\ee
is the Einstein tensor, the Ricci scalar $\bar R^\m{}_{\m}=\bar g^{\m\n}\bar R_{\m\n}$, and
\be\la{i7bb}
\bar T_{\a\b}=\le(\bar\epsilon+\bar p\ri)\bar u_\a\bar u_b+\bar p\bar g_{\a\b}\;,
\ee
is the energy-momentum tensor of matter governing the dynamic evolution of the background universe. In FLRW universe this tensor has the form of the perfect fluid with the background values of energy density  $\bar\epsilon$, pressure $\bar p$, and four-velocity $\bar u_\a$. The energy density
is defined in terms of the rest mass density, $\bar\rho$, and the internal energy per unit mass, $\bar\Pi$, as follows \citep{mtw}
\be\la{enrm1}
\bar\epsilon=\bar\rho \le(c^2+\bar\Pi\ri)\;,
\ee
where the internal energy density, $\bar\Pi$, is related to pressure, $\bar p$, by the law of conservation of energy \citep{mtw}
\be\la{enrm2}
d\bar\Pi+\bar pd\le(\frac{1}{\bar\rho}\ri)=0\;.
\ee

We assume that equations (\ref{i7})--(\ref{enrm2}) can be solved for a particular equation of state and, hence, the time evolution of the scale factor $a(\eta)$ is well-defined. Since all geometric and thermodynamic quantities are the implicit functions of $a(\eta)$, their temporal evolution is also well-defined.

\section{Post-Newtonian Approximations in Cosmology}\la{pnaco}

We assume that a localized distribution of matter (a {\it bare} perturbation) characterized by the tensor of energy-momentum $\mathfrak{T}_{\a\b}$ is placed on the background cosmological manifold. It makes a localized N-body astronomical system where the bodies interact to each other through their own gravitational field. The presence of the localized matter distribution in addition to the background matter of the Hubble flow, perturbs the background geometry as shown in (\ref{i2}) . Perturbed geometry is described by Einstein's field equations
\be\la{i7b}
G_{\a\b}=\frac{8\pi G}{c^4}\le( T_{\a\b}+{\cal T}_{\a\b}\ri)\;,
\ee
where $G_{\a\b}$, $T_{\a\b}$, and ${\cal T}_{\a\b}$ are the perturbed values of the Einstein tensor and the tensors of energy-momentum of the background matter and the localized system. The perturbations of the above tensors can be decomposed around their background values in a Taylor series with respect to the perturbation of the metric tensor $\varkappa_{\a\b}$,
\ba\la{i8}
G_{\a\b}&=&\bar G_{\a\b}+G^{(1)}_{\a\b}+G^{(2)}_{\a\b}+\ldots\;,\\\la{i9}
T_{\a\b}&=&\bar T_{\a\b}+T^{(1)}_{\a\b}+T^{(2)}_{\a\b}+\ldots\;\\\la{i10}
{\cal T}_{\a\b}&=&\phantom{\bar T_{\a\b}+}~{\mathfrak T}_{\a\b}+{\mathfrak T}^{(1)}_{\a\b}+\ldots\;,
\ea
where $G^{(1)}_{\a\b}$ and $T^{(1)}_{\a\b}$ are of the order of $O(\varkappa_{\a\b})$, $G^{(2)}_{\a\b}$ and $T^{(2)}_{\a\b}$ are of the order of $O(\varkappa^2_{\a\b})$, and so on. The tensor ${\mathfrak T}_{\a\b}$ is the source of the linear perturbations and, hence, is the same order of magnitude as $G^{(1)}_{\a\b}$ and $T^{(1)}_{\a\b}$. Perturbation ${\mathfrak T}^{(1)}_{\a\b}$ is due to the back gravitational reaction of the perturbed background geometry on the structure of ${\mathfrak T}_{\a\b}$.

After substituting expansions (\ref{i2}), (\ref{i8})-(\ref{i10}) to (\ref{i7b}), and accounting for the background Einstein equations (\ref{i7}), we obtain a set of partial differential equations for the metric tensor perturbation $\varkappa_{\a\b}$ which iterative solutions yield a successive post-Newtonian (sometimes called post-Friedmannian \citep{2002PhRvD..66j3507T}) approximations to the true geometry of the cosmological manifold \citep{koppetr}. Notice that the background geometry is well-defined by the solution of the Einstein equations (\ref{i7a}). We do not use any averaging procedure applied to (\ref{i7b}) in order to define the background manifold. This procedure is discussed, for example, in \citep{2010GReGr..42.1399K} and may be important in higher-order post-Newtonian approximations but it has no impact on the results of the present paper.

We focus on the Newtonian limit of the post-Newtonian approximations. The linear approximation is sufficient for this purpose. The field equations (\ref{i7b}) in this approximation read
\be\la{i11}
c^2 G^{(1)}_{\a\b}=\frac{8\pi G}{c^4}\le(T^{(1)}_{\a\b}+{\mathfrak T}_{\a\b}\ri)\;,
\ee
where
\be\la{i12}
G^{(1)}_{\a\b}\equiv\frac12\le(l_{\a\b}{}^{|\m}{}_{|\m} + \bar {g}_{\a\b} {B}^\m{}_{|\m} - {B}_{\a|\b} - {B}_{\b|\a}\ri)\;,
\ee
$T^{(1)}_{\a\b}$ is a rather complicated tensor expression depending on a particular cosmological equation of state \citep{koppetr}, the metric perturbation
\be\la{i13}
l_{\a\b}\equiv-\varkappa_{\a\b}+\frac12\bar g_{\a\b}\varkappa^\m{}_\m\;,
\ee
and vector $B^\a\equiv l^{\a\b}{}_{|\b}$.

It is well-known that an analytic coordinate transformation,
\be\la{i13a}
x'^\a=x^\a-\xi^\a(x^\b)\;,
\ee
induces a gauge transformation of the metric tensor perturbation \citep{1972gcpa.book.....W}
\be\la{i14}
l'_{\a\b}=l_{\a\b}-\xi_{\a|\b}-\xi_{\b|\a}+\bar g_{\a\b}\xi^\g{}_{|\g}\;.
\ee
Here, we keep only linear terms with respect to the gauge functions $\xi^\a$ assuming that these functions are of the same order of magnitude as $\varkappa_{\a\b}$. Higher-order terms can be included in (\ref{i14}) but this goes beyond the scope of the present paper (see \citep{1984CMaPh..94..379G,1988IJMPA...3.2651P}).

Calculation of the transformed Einstein tensor perturbation, $G'^{(1)}_{\a\b}$, defined by equation (\ref{i12}) along with perturbation (\ref{i14}), reveals that it is gauge-invariant in the sense that $G'^{(1)}_{\a\b}=G^{(1)}_{\a\b}$. Perturbation, $T^{(1)}_{\a\b}$, of the energy-momentum of the background matter is gauge-invariant, $T'^{(1)}_{\a\b}=T^{(1)}_{\a\b}$,  due to the background equations of motion of the background matter $\bar T^{\a\b}{}_{|\b}=0$ \citep{1988IJMPA...3.2651P,2005pfc..book.....M}. These gauge-invariant properties of tensors in the left side of (\ref{i12}) demand that the tensor of energy-momentum of the localized system obeyed the covariant law of conservation, ${\mathfrak T}^{\a\b}{}_{|\b}=0$. This equation can be reformulated in terms of partial derivatives \citep{1972gcpa.book.....W}
\be\la{i15}
\pd_\b\le(\sqrt{-\bar g}{\mathfrak T}^{\a\b}\ri)+\sqrt{-\bar g}\bar\G^\a{}_{\b\g}{\mathfrak T}^{\b\g}=0\;,
\ee
which is useful in defining the conserved mass, center of mass and a linear momentum of the localized astronomical system on the expanding FLRW manifold (see discussion in section \ref{nlfe}). We also notice that tensor ${\mathfrak T}^{\a\b}$ is gauge-invariant because it is the source of the linearized metric perturbation so that the gauge transformation (\ref{i14}) changes its form in terms of the second order that are not considered in the present paper.

\section{Post-Newtonian Field Equations in cosmology}\la{pnfec}

We are going to solve the Einstein field equations (\ref{i11}) in the linearized Hubble-parameter approximation which means that all terms of the order $\H^2$ will be discarded (these corrections can be found in \citep{koppetr}). This approximation is sufficient for discussing the effects of cosmological expansion in the solar system. Notice that the linearized Hubble approximation also discards all terms which are proportional to the time derivative of the Hubble parameter, $\H'$. This is because the background Einstein equations (\ref{i7}) suggests that $\H'\sim \H^2$ \citep{lnder,2005pfc..book.....M}.

By inspection, we can prove that perturbation, $T^{(1)}_{\a\b}=O(\H^2)$, that is it has no terms being linear with respect to $\H$. Indeed, the tensor of energy momentum $T_{\a\b}$ of the background cosmological matter is that of the ideal fluid
\be\la{bv3d}
T_{\a\b}=\le(\epsilon+ p\ri)u_\a u_b+ p g_{\a\b}\;,
\ee
which depends on the energy density $\epsilon$, pressure $p$, four-velocity of the matter $u^\a$, and the metric tensor $g_{\a\b}$. We expand tensor (\ref{bv3d}) in Taylor series around its background value, $\bar T_{\a\b}$, as shown in (\ref{i9}). Perturbation $T^{(1)}_{\a\b}$ is found by taking a variation of $T_{\a\b}$ from all variables. It yields,
\ba\la{asx1}
T^{(1)}_{\a\b}=\bar u_\a\bar u_\b\d\epsilon+\le(\bar u_\a\bar u_\b+\bar g_{\a\b}\ri)\d p+\le(\bar\e+\bar p\ri)\le(\bar u_\a\d u_\b+\bar u_\b \d u_\a\ri)+\bar p\d g_{\a\b}
\;,
\ea
where $\bar\epsilon$, $\bar p$, $\bar u^\a$, and $\bar g_{\a\b}$ are the background values of the energy density, pressure, four-velocity and the metric tensor respectively; $\d\epsilon$, $\d p$, $\d u_\a$ and $\d g_{\a\b}=\varkappa_{\a\b}$ are the perturbations of these variables, and we assume the Einstein summation rule. The terms in (\ref{asx1}) which are proportional to $\bar\e$ and $\bar p$, are of the order of $\H^2$ in virtue of the background Einstein equations (\ref{i6a})--(\ref{i7bb}). The perturbation $\d\e=(c^2/c_s^2)\bar\rho\d\mu$, where $c_s$ is the speed of sound of the background matter, $\bar\rho$ is the mass density of the background matter, and $\d\mu$ is the perturbation of the specific enthalpy, $\mu=(\e+p)/\rho$, of the background matter \citep{lnder,2005pfc..book.....M}. The $\bar\e\sim\bar\rho c^2\sim O(\H^2)$ due to the background Einstein equations. Finally, the second law of thermodynamics tells us that $\d p=\bar\rho\d\mu\sim O(\H^2)$. These estimates allow us to conclude that $T^{(1)}_{\a\b}\sim O(\H^2)$, and can be dropped out from equation (\ref{i11}) for we consider only terms of the linear order in $\H$.
It agrees with cosmological studies having been done by other researchers \cite{2005pfc..book.....M,2010deto.book.....A} (see \citep{koppetr} for more detail).

Equation (\ref{i11}) becomes
\be\la{i16}
l_{\a\b}{}^{|\m}{}_{|\m} + \bar {g}_{\a\b} {B}^\m{}_{|\m} - {B}_{\a|\b} - {B}_{\b|\a}=\frac{16\pi G}{c^4}{\mathfrak T}_{\a\b}\;,
\ee
where $B^\a\equiv l^{\a\b}{}_{|\b}$.
It is gauge invariant and admits a gauge freedom (\ref{i14}) in choosing the metric perturbations, $l_{\a\b}$, on the background manifold. The gauge freedom can be used in order to simplify solution of (\ref{i16}). It is convenient to fix the gauge by imposing the following gauge condition \citep{2001PhLA..292..173K,2002PhLB..532....1R}
\be
\la{qe6}
B^\a=-\frac{2\H}{c} l^{\a\b}\bar u_\b\;,
\ee
which generalizes the harmonic (deDonder) gauge condition applied in the post-Newtonian approximations performed on the asymptotically-flat background \citep{2011rcms.book.....K,1987thyg.book..128D}, to cosmology. To understand the advantage of the gauge (\ref{qe6}) in the cosmological setting, we write down the covariant Laplace-Beltrami operator for the metric perturbation in terms of the partial derivatives as follows
\be
\la{qe7}
l_{\a\b}{}^{|\m}{}_{|\m}=\frac1{a^2}\Box l_{\a\b}+\frac{2{\cal H}}{c}\bar u^\m \pd_\m l_{\a\b}-\frac2c\le({\cal H}\bar u^\m l_{\m\a}\ri)_{|\b}
-\frac2c\le({\cal H}\bar u^\m l_{\m\b}\ri)_{|\a}+O\le(\frac{\H^2}{c^2}\ri)\;,
\ee
where
\be\la{ge7a}
\Box\equiv\mm^{\m\n}\frac{\pd^2}{\pd x^\m\pd x^\n}=-\frac1{c^2}\frac{\pd^2}{\pd\eta^2}+\frac{\pd^2}{\pd x^2}+\frac{\pd^2}{\pd y^2}+\frac{\pd^2}{\pd z^2}\;,
\ee
is the wave operator in Minkowski spacetime, and we have neglected all terms of the order of $\H^2$. Substituting (\ref{qe6}) and (\ref{qe7}) to (\ref{i16}) cancels out a significant number of terms, thus, reducing equation (\ref{i16}) to a much simpler form
\be\la{i17}
\Box l_{\a\b}+\frac{2{\cal H}}{c}a^2\bar u^\m\pd_\m l_{\a\b}=\frac{16\pi G}{c^4}a^2 \mathfrak{T}_{\a\b}\;.
\ee
This equation can be reduced to the wave equation by making use of the following steps.

We define the conformal metric perturbations, $\gamma_{\a\b}$ and $h_{\a\b}$, by equations
\ba\la{i18}
g_{\a\b}&=&a^2(\eta)\le(\mm_{\a\b}+h_{\a\b}\ri)\;,\\\la{i18a}
l_{\a\b}&=&a^2(\eta)\gamma_{\a\b}\;,
\ea
so that
\be\la{i19}
h_{\a\b}=-\gamma_{\a\b}+\frac12\mm_{\a\b}\g^\m{}_\m\;,
\ee
and indices of $h_{\a\b}$ and $\g_{\a\b}$ are raised and lowered with the metric $\mm_{\a\b}$. Substituting (\ref{i18a}) to (\ref{i17}) and simplifying terms, yield
\be\la{as1}
\Box\g_{\a\b}-\frac{2H}{c^2}\pd_\eta\g_{\a\b}=\frac{16\pi G}{c^4}\mathfrak{T}_{\a\b}\;,
\ee
where the conformal Hubble parameter $H=\dot a/a=\H'/a$ and all terms of the order of $H^2$ have been discarded. Equation (\ref{as1}) explicitly contains a term depending on the Hubble parameter but it can be eliminated by introducing a new variable, $a\g_{\a\b}$. In terms of this variable, equation (\ref{as1}) becomes equivalent to
\be\la{as2}
\Box\le(a\g_{\a\b}\ri)=\frac{16\pi G}{c^4} a\mathfrak{T}_{\a\b}\;.
\ee
This is a wave equation in conformally-flat spacetime which can be easily solved with the standard technique of retarded potential if we impose the boundary condition of vanishing $a\g_{\a\b}$ and $a\g_{\a\b,\m}$ at conformal past-null infinity \citep{ZAMM19650450133}. The advanced potential violates causality at higher post-Newtonian approximations, and we do not consider it here (see \citep{2007arXiv0705.0083I} for further discussion). The retarded solution of (\ref{as2}) is
\be\la{as3}
a(\eta)\g_{\a\b}(\eta,{\bm x})=-\frac{4G}{c^4}\int_{\cal V}\frac{a(s)\mathfrak{T}_{\a\b}\le(s,{\bm x}'\ri)d^3x'}{|{\bm x}-{\bm x}'|}\;,
\ee
where ${\cal V}$ is the volume of integration occupied by matter, $\mathfrak{T}_{\a\b}$, of the localized astronomical system, and the retarded time $s$ is defined on the null cone of Minkowski spacetime,
\be\la{rtime}
s=\eta-\frac1c|{\bm x}-{\bm x}'|\;.
\ee
Solution (\ref{as3}), (\ref{rtime}) is given in conformal coordinates  $x^\a=\{c\eta,x^i\}$. It is Lorentz-invariant with respect to the Lorentz transformations of the conformal coordinates and satisfies the gauge condition (\ref{qe6}).

It is important to emphasize that the retarded metric potentials (\ref{as3}) describe gravitational field everywhere both in the near zone and in the radiative zone of the localized astronomical system emitting gravitational waves. Propagation of gravitational waves in the expanding universe takes place on the hypersurface of a null cone (\ref{rtime}) in the conformal spacetime. Therefore, the speed of weak gravitational waves is equal to the speed of light, $c$, referred to the conformal coordinates $x^\a=\{c\eta,{\bm x}\}$.

Solution of (\ref{as2}) is defined up to the residual gauge transformation (\ref{i13a}) with the gauge functions $\xi^a$ obeying the homogeneous wave equation
\be\la{hwe1}
\Box (a\xi^a)=0\;.
\ee
This equation is valid up to the terms of the linear order of $O(\H)$. Quadratic corrections of the order of $O(\H^2)$ to equation (\ref{hwe1}) are given in \citep{koppetr}.

\section{The Newtonian Limit of the Field Equations}\la{nlfe}

Equation (\ref{as3}) yields the post-Newtonian potentials $\g_{\a\b}$ in the linearized approximation with respect to the Hubble parameter $\H$ which is incorporated implicitly to the solution through the time-dependent conformal factor $a=a(\eta)$. Only the potential $\g_{00}$ is relevant in discussion of the Newtonian-theory approximation which is defined as a limit of the retarded solution (\ref{as3}) when the parameter $1/c\rightarrow 0$.

The Newtonian potential, $U$, in expanding universe is defined as $U=-c^2\gamma_{00}/4$. In what follows, we shall focus on the localized system consisting of one massive body. Our treatment can be easily extended to N-body problem for, in the approximation under consideration, the Newtonian potential obeys the principle of a linear superposition. Equation (\ref{as3}) yields
\be\la{geo17}
U(\eta,{\bm x})=\frac{G}{c^2}\int_{{\cal V}}\frac{a(s)\mathfrak{T}_{00}\le(s,{\bm x}'\ri)d^3x'}{a(\eta)|{\bm x}-{\bm x}'|}\;,
\ee
where the integration is performed over the volume of the massive body. We assume the matter inside the body moves slowly, $\dot x^i\ll c$. Hence, we can expand the retarded function in the integrand of (\ref{geo17}) in the Taylor series around the present time $\eta$. It leads to
\be\la{geo19}
U(\eta,{\bm x})=\frac{G}{c^2}\int_{{\cal V}}\frac{\mathfrak{T}_{00}\le(\eta,{\bm x}'\ri)d^3x'}{|{\bm x}-{\bm x}'|}+\frac{G}{c^3a(\eta)}\int_{{\cal V}}\frac{\pd}{\pd \eta}\le[a(\eta)\mathfrak{T}_{00}\le(\eta,{\bm x}\ri)\ri]d^3x+O\le(\frac{G}{c^4}\ri)\;.
\ee
The first term in right side of (\ref{geo19}) looks exactly as the Newtonian potential of a conventional Newtonian theory. Yet, we cannot use this expression for doing the multipolar expansion with respect to the gravitational harmonics - we still need to define the concept of the conserved mass, dipole moment and other multipoles of the astronomical system that will be independent of the Hubble parameter. Moreover, the correct Newtonian limit demands that the second integral in the right side of (\ref{geo19}) would be, at least, of the second order of magnitude with respect to $\H$ which we systematically discard.

These issues can be resolved with the help of the covariant law of conservation of the tensor of energy-momentum $\mathfrak{T}_{\a\b}$ - equation (\ref{i15}). Making use of (\ref{i6}) for the Christoffel symbols  results in a more explicit form of (\ref{i15}),
\ba\la{geo20}
\frac1c\frac{\pd}{\pd \eta}\le(a\mathfrak{T}_{00}\ri)-\frac{\pd}{\pd x^j}\le(a\mathfrak{T}_{0j}\ri)&=&-\frac Hc a\mathfrak{T}_{kk}\;,\\\la{geo21}
\frac1c\frac{\pd}{\pd \eta}\le(a\mathfrak{T}_{i0}\ri)-\frac{\pd}{\pd x^j}\le(a\mathfrak{T}_{ij}\ri)&=&-\frac Hc a\mathfrak{T}_{i0}\;,
\ea
Implementing (\ref{geo20}) for calculating the second integral in (\ref{geo19}), yields
\be\la{gek}
\int_{{\cal V}}\frac{\pd}{\pd \eta}\le(a\mathfrak{T}_{00}\ri)d^3x=
-aH\int_{{\cal V}}\mathfrak{T}_{kk}d^3x\;,
\ee
where we have applied the Gauss theorem to eliminate the integral for the divergence of the vector field $a\mathfrak{T}_{0j}$.
Further application of the laws of conservations (\ref{geo20}), (\ref{geo21}) brings about the virial relationships
\ba\la{geo22}
\int_{{\cal V}}a\mathfrak{T}_{ij}d^3x=\frac1{2c^2}\frac{d^2}{d \eta^2}\int_{{\cal V}}a\mathfrak{T}_{00}x^i x^j d^3x+O(\H)\;,\\
\la{geo23}
\int_{{\cal V}}a\mathfrak{T}_{i0}d^3x=-\frac1c\frac{d}{d \eta}\int_{{\cal V}}a\mathfrak{T}_{00}x^i  d^3x+O(\H)\;,
\ea
where $O(\H)$ indicates the residual terms of the order of the Hubble parameter $\H$.

Let us assume that distribution and motion of matter inside the localized system, is stationary. Relationships (\ref{geo22}), (\ref{geo23}) suggest, then, that the right side of equation (\ref{gek}) is of the order of $\H^2$, and can be neglected in the linearized Hubble approximation. This result also suggests that we can define the conserved mass, $M$, of the massive body residing in the expanding universe by the following integral
\be\la{geo24}
M=\frac1{c^2}\int_{{\cal V}}a(\eta)\mathfrak{T}_{00}(\eta,{\bm x})  d^3x\;.
\ee
The mass is conserved, $dM/d\eta=0$, up to the terms of the second order in $\H^2$. Since $d/d\eta=a(\eta)d/dt$, the law of conservation of mass can be recast to
\be\la{geo25}
\frac{dM}{dt}=0\;,
\ee
where we have used the cosmological time $t$ that coincides with the proper time measured by the Hubble observer.

We define the dipole moment, $I^i$, and the linear momentum, $P^i$, of the massive body in the expanding universe by
\ba\la{i20}
I^i&=&\frac1{c^2}\int_{{\cal V}}a(\eta)\mathfrak{T}_{00}(\eta,{\bm x})x^i  d^3x\;,\\
P^i&=&-\frac1c\int_{{\cal V}}a^2(\eta)\mathfrak{T}_{i0}(\eta,{\bm x})  d^3x\;.
\ea
Equation (\ref{geo23}) yields a relationship between a time derivative of $I^i$ and $P^i$,
\be\la{i21}
\frac{dI^i}{dt}=P^i\;.
\ee
Taking a time derivative from $P^i$ and applying (\ref{geo21}) along with (\ref{geo23}) yields the law of conservation of a linear momentum of the isolated self-gravitating body in the expanding universe
\be\la{i22}
\frac{dP^i}{dt}=0\;.
\ee
Equations (\ref{i21}), (\ref{i22}) tell us that we can always chose the inertial coordinates in the expanding universe such that the center of mass of the body is defined by the condition, $I^i=0$, and it remains at rest or moves with a constant velocity in these coordinates. After choosing the center of mass at the origin of the inertial coordinates, $x^\a=\{c\eta,x^i\}$, the multipolar expansion of the Newtonian potential (\ref{geo19}) reads
\be\la{huw1}
U(\eta,{\bm x})=\frac{G}{a(\eta)}\frac{M}{|{\bm x}|}+\ldots\;,
\ee
where ellipses denote quadrupole, octupole, and higher-order harmonics, and $|{\bm x}|$ is the radial distance from the spatial origin of the conformal coordinates to the field point. The multipolar expansion (\ref{huw1}) is valid in the linearized Hubble approximation with all terms of the order of $\H^2$ having been abandoned.

The Newtonian potential (\ref{huw1}) looks like the potential of a point-like mass $M$ with the effective gravitational constant $\mathfrak{G}(\eta)\equiv G/a(\eta)$, depending on time. It may lead to a conclusion that the expansion of the universe encompasses a secular decrease of the gravitational constant - the idea pioneered by P.M. Dirac \citep{Dirac05041938} and reiterated by R. Dicke in his weak anthropic principle \citep{1961Natur.192..440D} and a scalar-tensor theory of gravity (along with C. Brans) \citep{1961PhRv..124..925B}. This conclusion is too straightforward and is invalid in general relativity. The reason is that the gravitational constant can be measured only through observing orbital motion of celestial bodies and processing the observational data on the basis of the Newtonian equations of motion. Scrutiny analysis given in next section shows that the scale factor $a(\eta)$ drops out of the equations of motion after making a transformation from the conformal to local coordinates, so that the effective gravitational constant $\mathfrak{G}(\eta)$ gets rescaled and its variation cannot be observed -- only the constant value of $G$ can be measured.

\section{The Newtonian equations of motion in expanding universe}\la{nemeu}

Astronomical observations make use of test particles of two sorts: with a non-zero rest mass and massless photons (light, radio). In order to understand the impact of cosmological expansion on the motion of particles we have to derive relativistic equations of their motion and to analyze their Newtonian limit which is reached for the value of the parameter $1/c\rightarrow 0$. The task is to find out the analytic dependence of the Newtonian equations of motion on the Hubble parameter $\H$.

Any test particle, if it is unaffected by an external force of non-gravitational origin, moves along a geodesic worldline which equation is given by
\be\la{geo1}
\frac{d^2x^\a}{d\t^2}+\G^\a{}_{\m\n}\frac{dx^\m}{d\t}\frac{dx^\n}{d\t}=0\;,
\ee
where $\t$ is an affine parameter on the world line of the particle, $\G^\a{}_{\b\g}$ is the perturbed affine connection given by
\be\la{geo2}
\G^\a{}_{\m\n}=\frac12 g^{\a\b}\le(g_{\b\m,\n}+g_{\b\n,\m}-g_{\m\n,\b}\ri)\;,
\ee
and the perturbed metric $g_{\a\b}$ is defined by (\ref{i18}) with $h_{\a\b}$ taken from (\ref{i19}) and (\ref{as3}). The affine parameters for time-like and light-like geodesics are different \citep{1972gcpa.book.....W}. Irrespectively of the nature of the affine parameter we can always choose the coordinate time $\eta$ as an independent parameter along geodesics. Equation (\ref{geo1}) parameterized by the coordinate time $\eta$, reads
\be\la{geo7}
\ddot x^i+\G^i{}_{\m\n}\dot x^\m\dot x^\n-\frac1c\G^0{}_{\m\n}\dot x^\m\dot x^\n\dot x^i=0\;.
\ee
We elaborate on these equations in the next two subsections. We shall focus primarily on discussion of the Newtonian limit of these equations in application to test particles with a rest mass, and to light (massless particle, photon).

\subsection{Equations of motion of test particles with non-zero rest mass}\la{hjqcv}

Calculating the Christoffel symbols (\ref{geo2}) and substituting them to (\ref{geo7}), yield the equations of motion of test particles
\be\la{hus}
\ddot{\bm x}=-H\dot{\bm x}-\frac{GM}{a|{\bm x}|^3}{\bm x}\;,
\ee
where all terms of the post-Newtonian order $\sim 1/c^2$ and the order of $H^2$, have been omitted.
These equations are written in the conformal coordinates $\{\eta,{\bm x}\}$ and apparently have terms which makes them looking unlike the Newtonian equations of classical mechanics. In particular, we notice the presence of the conformal Hubble parameter $H$ and the time-dependent scale factor $a(\eta)$ that again makes an impression that the effective gravitational constant $\mathfrak{G}(\eta)=G/a(\eta)$ appears in the equations.

It is remarkable that there are transformations of the conformal coordinates which convert (\ref{hus}) to exactly Newtonian form of the classic mechanics. The coordinate transformations are
\be\la{hus1}
t=\int^\eta_{\eta_0}a(\eta')d\eta'\;,\qquad\quad {\bm r}=a(\eta){\bm x}\;,
\ee
where we integrate from the initial epoch, $\eta_0$, and assume that at the initial epoch the dimensionless conformal factor $a(\eta_0)=1$.
These transformations reduce equation (\ref{hus}) to the exact form of the classic Newtonian equations of motion
\be\la{hus2}
{\bm r}''=-\frac{GM}{r^3}{\bm r}\;,
\ee
where $r=|{\bm r}|$, and a prime denotes a derivative with respect to time $t$.

The obtained equation is fully identical with the Newtonian equations of motion in the classic celestial mechanics, and does not predict any ``cosmological force'' that would be proportional to the Hubble constant $\H$. All terms depending on $H$ canceled out after applying the conformal space and time transformation (\ref{hus1}). The explicit dependence of the Newtonian force on the scale factor, $a(\eta)$, has disappeared as well. Our derivation of equation (\ref{hus2}) helps to understand the result obtained by Brumberg and Krasinsky \citep{2004CeMDA..90..267K} with a different mathematical technique. They analyzed circular equations of motion by solving equations (\ref{hus}) explicitly, and arrived at the conclusion that the circular orbits are not affected by the cosmological expansion. In fact, this is a consequence of the conformal transformation (\ref{hus1}) which allows us to prove that in FLRW universe any massive particle moves in accordance with the Newtonian equations of motion without any presence of the Hubble parameter $H$. More advanced post-Newtonian analysis of the residual terms in the Newtonian equations of motion show that the residual terms are proportional to $\sim H\dot x^2/c$ and $\sim H^2$. Hence, any Keplerian orbit of massive particle (not only circular) remains unaffected by the cosmological expansion. This conclusion has been validated by other researchers who used different mathematical techniques \citep{2000CQGra..17.2739B,2007CQGra..24.5031M,2010RvMP...82..169C}.

\subsection{Equations of motion of light}\la{emli}

Light is a wave of electromagnetic field which obeys homogeneous Maxwell's equations in free space. These equations are conformally-invariant as we prove in the next section. For these reasons, the equation of motion of light in the conformal coordinates $\{\eta,x^i\}$ is the same as in Minkowskian spacetime (remember that we consider the case of FLRW universe with space curvature $k=0$)
\be\la{hus3}
\ddot{\bm x}=0\;.
\ee
Hereby, we have neglected the gravitational force exerted on light by the localized astronomical system as it produces the deflection of light and time delay which are of the post-Newtonian order and are not of the primary concern in the present paper. We draw attention of the reader that equation (\ref{hus3}) does not have any residual term depending on the Hubble parameter $\H$. This is because of the conformal invariance of the Maxwell equations \citep{wald}.

Equation (\ref{hus3}) can be also derived from (\ref{geo7}) by calculating the Christoffel symbols, making use of equation of light cone, $\mm_{\a\b}\dot x^\a\dot x^\b=0$, and neglecting the gravitational deflection of light caused by the localized astronomical system. Equation (\ref{hus3}) tells us that light in conformal coordinates of the conformally-flat FLRW manifold propagates with the constant speed $c$ along straight lines. Equation (\ref{hus3}) looks exactly the same as the Newtonian equation of propagation of light but one has to use this analogy with care. The point is that the conformal time, $\eta$, and space coordinates, ${\bm x}$, entering (\ref{hus3}), are not those $\{t,{\bm r}\}$ used in the Newtonian equations of motion (\ref{hus2}) of massive particles. Therefore, we can not use solution (\ref{hus3}) to analyze the orbital motion of massive particles.

In order to make equations (\ref{hus3}) compatible with the Newtonian equations of motion (\ref{hus2}) we should make use of the conformal coordinate transformations (\ref{hus1}) in equation (\ref{hus3}). It recasts (\ref{hus3}) to
\be\la{hus4}
{\bm r}''=\H{\bm r}'\;.
\ee
This equation should be compared with (\ref{z3}) that we would expect in the Newtonian limit of general relativity in coordinates, $\{t,{\bm r}\}$. However, the conformal nature of the cosmological spacetime and the conformal invariance of Maxwell's equations does not support this expectation and bring about equation (\ref{hus4}) that yields the correct Newtonian limit for the equation of propagation of light in cosmology with the space curvature $k=0$.
Similarly to (\ref{z3}), equation (\ref{hus4}) describes propagation of photons along straight lines but we notice the presence of a force, $\H{\bm r}'$, in its right side that is acting on photon due to the expansion of the universe. Appearance of this force points out that coordinates, $\{t,{\bm r}\}$, can not be considered as inertial coordinates for description of light propagation.

Equation (\ref{hus4}) can be solved exactly but it will introduce terms of the order of $\H^2$ and higher, which we neglect. We are interested only in keeping linear terms with respect to $\H$. For this reason, we can assume that $\H$ in the right side of (\ref{hus4}), is constant. It allows us to directly integrate this equation with respect to time. The integration yields a {\it coordinate} velocity of light,
\be\la{hus4a}
{\bm r}'=(1+\H t)c{\bm k}\;,
\ee
where the unit vector ${\bm k}=\{k^i\}$ is directed from the point of emission of light toward the point of its reception $(\d_{ij}k^ik^j=1)$. The non-inertial force $\H{\bm r}'$ accelerates photons with respect to spatial coordinates, ${\bm r}$, so that the {\it coordinate} speed of light, ${\bm r}'$, does not remain $c$ but grows directly proportional to the cosmological time $t$. We emphasize that there is no contradiction with special relativity because the physical speed of light, measured locally at time $t=0$, remains equal to the fundamental speed $c$. One more integration of (\ref{hus4a}) yields trajectory of the light ray in coordinates $\{t,{\bm r}\}$,
\be\la{hus5}
{\bm r}={\bm\xi}+c{\bm k}\le(t+\frac12\H t^2\ri)\;,
\ee
where ${\bm\xi}=\{\xi^i\}$ is the (constant) impact parameter of the light ray with respect to the origin of the spatial coordinates. The impact parameter is orthogonal to ${\bm k}$ in the sense that the Euclidean dot product of the two vectors vanishes, $\d_{ij}k^i\xi^j=0$ (more details about the geometric properties of the parametrization (\ref{hus5}) see in our textbook \citep[section 7.3]{2011rcms.book.....K}).

Another way to understand the propagation of light in the expanding universe is to apply a slightly different conformal transformation of time and space coordinates, namely,
\be\la{hus6}
\lambda=a(\eta)(\eta-\eta_0)\;,\qquad\quad {\bm r}=a(\eta){\bm x}\;,
\ee
where $\lambda$ is a new time coordinate which differs from the conformal time $\eta$ by the time-dependent scale factor $a(\eta)$.
This transformation leaves equation (\ref{hus3}) form-invariant, in the sense that if all terms of the order of $\H^2$ are neglected, the equation of light propagation in coordinates $\{\lambda,{\bm r}\}$ has the same form as in the conformal coordinates $\{\eta,{\bm x}\}$,
\be\la{hus7}
\frac{d^2{\bm r}}{d\lambda^2}=0\;.
\ee
Equation (\ref{hus7}) means that the coordinates $\{\lambda, {\bm r}\}$ can be considered as inertial coordinates for describing propagation of light.

Solution of equation (\ref{hus7}) is a straight line
\be\la{hus8}
{\bm r}={\bm\xi}+c{\bm k}\lambda\;,
\ee
with the constant {\it coordinate} speed of light, $d{\bm r}/d\lambda=c{\bm k}$, and where ${\bm k}$ and ${\bm\xi}$ have the same meaning as in (\ref{hus5}). By expanding the conformal factor in (\ref{hus6}) in a  Taylor series around $\eta=\eta_0$ and making use of (\ref{hus1}), we get a relationship between the two time scales,
\be\la{hus9}
\lambda=t+\frac12\H t^2\;,
\ee
that matches equation (\ref{hus8}) with equation (\ref{hus5}).

The lesson we have learned from the analysis of the equations of motion in the conformal FLRW spacetime, is that there is no single inertial coordinate chart in the expanding universe in the Newtonian description of motion of massive particles and light. Transformation to inertial coordinates, $\{t,{\bm r}\}$, eliminate all Hubble-dependent terms of the order of $\H$ from the Newtonian equations of motion of massive particles but it can not eliminate such terms from the Newtonian equation of motion of light. On the other hand, transformation to the inertial coordinates $\{\lambda,{\bm r}\}$ eliminate all Hubble-dependent terms of the order of $\H$ from the equation of motion of light but put them back to the equations of motion of massive particles.

There is a physical difference between the two time scales, $t$ and $\lambda$. The cosmological time $t$ coincides with the proper time of the Hubble observers while the time scale $\lambda$ diverges quadratically from $t$ and does not coincide with the proper time. Special relativity postulates that proper time can be measured with an ideal atomic clock. This postulate should be checked in cosmological spacetime to make sure that atomic clocks do measure the proper time.

\section{Relationship between atomic and proper time in cosmology}\la{atpr}

General relativity postulates that the proper time of observer is measured by an {\it ideal} clock carried by the observer.  The most precise practical realization of the {\it ideal} clock is the atomic frequency standard which operational principle is based on the use of electron's energy level transitions in atom to produce a stable frequency that drives the clock \citep{lomhejeff}. In classic regime the atomic frequency is defined by the difference between the energies of two orbits of an electron moving around nucleus of the atom in accordance with the second Newton's law with the central Coulomb force. Additional Bohr's postulate is required in order to match the classic motion of electron in atom with quantum mechanics but it is not essential for further consideration. Thus, in order to make sure that the proper and atomic time scales in the expanding universe coincide, we have to prove that the orbital motion of an electron in a hydrogen atom is not affected by the expansion of the universe, at least, in the linearized Hubble approximation. The problem was studied in \citep{1964PhRvL..12..435D,1999CQGra..16.1313B,2010RvMP...82..169C} and the authors have reached a consensus that the expansion of the universe should not affect the stability of atomic frequencies. Here, we refine the proof of this statement with a more rigorous mathematical technique.

Electromagnetic field tensor $F_{\a\b}=\pd_\a A_{\b}-\pd_\b A_{\a}$ is defined in terms of a vector potential $A_\a$. General-relativistic
Maxwell's equation for electromagnetic vector $A^\a$ in a curved FLRW spacetime are \citep{mtw}.
\be\la{at1}
A_\a{}^{|\b}{}_{|\b}-A^\b{}_{|\b\a}-\bar R_{\a\b}A^\b=-\frac{4\pi}{c}J_\a\;,
\ee
where $J_\a$ is a four-vector of electric current. Usually, the Lorentz gauge condition, $A^{\b}{}_{|\b}=0$, is imposed on vector-potential in order to eliminate the middle (gauge-dependent) term in the left side of equation (\ref{at1}) and reduce it to the de Rham equation \citep{mtw}. Unfortunately, such a gauge condition does not simplify the last term in the left side of (\ref{at1}) depending on the Ricci tensor. We have discovered that another gauge condition is more elegant (compare with (\ref{qe6}) for gravitational case),
\be\la{at2}
A^\b{}_{|\b}=-\frac{2\H}{c} A^\b\bar u_\b\;.
\ee
This gauge condition has been also found by M. Ibison \citep{ibison:28,ibison2011} in his study of electrodynamics in cosmology with a future conformal horizon.

We substitute (\ref{at2}) into the second term (\ref{at1}) and use expression (\ref{i6a}) for the Ricci tensor. It turns out that after performing the covariant derivatives and reducing similar terms, all the terms depending on the Hubble parameter cancel out in all orders, and we arrive to {\it exact} Maxwell's equation in FLRW spacetime (with space curvature $k=0$),
\be\la{at3}
\Box A_\a=-\frac{4\pi}{c}a^2J_\a\;,
\ee
where $\Box$ is the wave operator in the Minkowski spacetime defined in (\ref{ge7a}).
It is remarkable that it look similar to special relativity in Minkowski space-time except for the scale factor, $a=a(\eta)$, that appears in the right side of (\ref{at3}). The reason for the simplicity of (\ref{at3}) is the conformal invariance of electromagnetic field which makes the left side of (\ref{at1}) be equal to the left side of (\ref{at3}) up to the conformal factor $1/a^2$ \citep{wald}. We take a physically-relevant retarded solution of equation (\ref{at3}),
\be\la{at3a}
A_\a(\eta,{\bm x})=\frac1c\int_{\st V}\frac{a^2(s)J_a(s,{\bm x}')d^3x'}{|{\bm x}-{\bm x}'|}\;,
\ee
where the retarded time $s$ is defined in (\ref{rtime}), and the volume integral is performed over the charge distribution inside the atomic nucleus. Because the retarded argument $s$ in the solutions for electromagnetic and gravitational field is defined by the same equation (\ref{rtime}), weak electromagnetic and gravitational waves propagate with the same speed $c$ in the conformally-flat space of FLRW universe.

We are interested in the Coulomb approximation for the potential $A_\a$ to describe the electric field produced by atom's nucleus. In this approximation, it is sufficient to consider only the electric potential $\phi\equiv cA_0$.
Everywhere inside the near-zone of the atom, the retarded argument of the vector potential can be expanded in a Taylor series around time $\eta$. For the electric potential it yields
\be\la{at3b}
\phi(\eta,{\bm x})=\int_{\st V}\frac{a^2(\eta)J_0(\eta,{\bm x}')d^3x'}{|{\bm x}-{\bm x}'|}+\frac1{c}\int_{\st V}\frac{\pd}{\pd\eta}\le[a^2(\eta)J_0(\eta,{\bm x}')\ri]d^3x'+O\le(\frac1{c^2}\ri)\;.
\ee
Taking covariant divergence from both sides of equation (\ref{at1}) reveals that the divergence of the left side vanishes identically. It means that the electric current is conserved,
\be\la{at4}
J^\a{}_{|\a}=0\;,
\ee
which is equivalent to
\be\la{at5}
\frac1c\frac{\pd}{\pd \eta}\le(a^2 J_{0}\ri)-\frac{\pd}{\pd x^i}\le(a^2 J_{i}\ri)=0\;.
\ee
The total electric charge $Q$ is defined by
\be\la{at6}
Q=-\int_{\st V}a^2(\eta)J_0(\eta,{\bm x})d^3x\;.
\ee
The charge is conserved, $dQ/d\eta=0$, up to terms of the order of $\H^2$  as follows from (\ref{at5}). Replacing the derivative with respect to the conformal time $\eta$ by the derivative with respect to cosmological time, the law of conservation of charge in expanding universe becomes
\be\la{at6a}
dQ/dt=0\;.
\ee
Because of charge's conservation, the second term in right side of (\ref{at3b}) vanishes while the first term can be expanded in terms of the electric multipoles -- the constant charge $Q$, the dipole electric moment $Q_i$, and so on,
\be\la{at7}
\phi(\eta,{\bm x})=-\frac{Q}{|{\bm x}|}+\frac{Q_i x^i}{|{\bm x}|^3}+...\;.
\ee
In what follows, we neglect the dipole and higher-order multipole moments of electric field.

The atomic frequency of electromagnetic wave emitted by atom, is defined by the difference between two values of orbital energies of an electron when it jumps from one orbit of the atom to another. The orbital motion of an electron in a curved spacetime is given by the second Newton's law with an electromagnetic Lorentz force in its right side \citep{mtw} while the left side of this law is a covariant derivative from the linear momentum of electron $p^\a={m}dx^\a/d\tau$,
\be\la{at8}
{m}\le(\frac{d^2x^a}{d\t^2}+\G^\a{}_{\b\g}\frac{dx^\b}{d\t}\frac{dx^\g}{d\t}\ri)=eF^\a{}_{\b}\frac{dx^\b}{d\t}\;,
\ee
where $\t$ is the affine parameter along the electron's orbit, $e$ is the charge of electron $(e<0)$, and ${m}$ is electron's mass. We accept that both charge and mass of electron remain constant in expanding universe in accordance with the definitions of these conserved quantities given in (\ref{at6}) and (\ref{geo24}). The Christoffel symbols in (\ref{at8}) account for the presence of gravitational field. It is sufficient to take into account only the background value of the Christoffel symbols, $\bar\G^\a{}_{\m\n}$, as the metric tensor perturbations cause negligible effects at least, in the solar system.  Strong gravitational field of a localized gravitational system (a black hole) may affect the clock's performance \citep{2004PhRvL..93x0401G} but we do not discuss the strong field effects in the present paper.

If the conformal time, $\eta$, is used for parametrization of electron's orbit, equation of motion (\ref{at8}) assumes the following form
\be\la{at9}
\ddot x^i+\bar\G^i{}_{\m\n}\dot x^\m\dot x^\n-\frac1c\bar\G^0{}_{\m\n}\dot x^\m\dot x^\n\dot x^i=\frac{e}{{m}}\le(F^{i}{}_\b-\frac1c F^0{}_\b\dot x^i\ri)\dot x^\b\dot\t\;.
\ee
In the slow-motion approximation $\dot\t=a(\eta)$, $F^i{}_0=F_{i0}/a^2$, and the electric field of the atomic nucleus is $E_i=cF_{i0}=\pd_i\phi$. By neglecting relativistic corrections of the order of $1/c^2$, we obtain the equation of motion of electron in the expanding universe
\be\la{at10}
\ddot{\bm x}=-H\dot{\bm x}+\frac{eQ}{{m}a}\frac{{\bm x}}{|{\bm x}|^3}\;,
\ee
which mathematical structure is similar to the Newtonian equation of motion of massive test particles (\ref{hus}). Hence, by doing conformal transformation (\ref{hus1}) we can reduce (\ref{at10}) to the classic form of the second Newton's law with Coulomb's force
\be\la{at11}
{m}{\bm r}''=\frac{eQ}{r^3}{\bm r}\;.
\ee
Thus, choosing coordinates $\{t,{\bm r}\}$ on the expanding cosmological background eliminates any dependence of the orbital motion of electrons on the scale factor and the Hubble parameter. This proves that the atomic frequencies are not affected by the expansion of universe, at least, up to terms of the order of $\H^2$. Furthermore, the time parametrization of equation (\ref{at11}) makes it evident that the atomic time (AT), measured by the ideal atomic clock carrying out by a Hubble observer, is identical with the proper time, $\t$, of the observer as well as with the cosmological time $t$.

Comparing two equations, (\ref{hus2}) and (\ref{at11}), we observe that the parametrization of the orbital motion of massive particles in gravitational field and that of electrons in atoms is done with the same parameter -- the cosmological time $t$. The argument of the gravitational equations of motion of massive particles (\ref{hus2}) is called the ephemeris time (ET). Our consideration makes it clear that in general relativity ET can be measured in two different ways: (1) by observing motion of massive particles in gravitational field, and (2) making use of atomic clocks. If general relativity is valid we should not see any divergence between the two time scales -- ET and AT.

In contrast to ET and AT time scales, the time scale, $\lambda$, that parametrizes the equations of motion of photons in expanding universe, diverges quadratically from the atomic time (AT) in accordance with equation (\ref{hus9}). This divergence has important physical consequences in astrophysical applications and in testing general relativity as it allows us to measure the local value of the Hubble parameter $\H$ independently of cosmological observations.

\section{The impact on Astronomical Ephemerides}
\epigraph{Modern ephemerides represent possibly the greatest dynamical system of all time...}{Myles. E. Standish,  \citep{2005tvnv.conf..163S}}

\subsection{What equations should we utilize for space navigation?}

Astronomical observatories and the national Space Navigation Centers (like NASA JPL) operate with the computer codes calculating astronomical ephemerides of celestial bodies under assumption that the Newtonian equations (\ref{z1})--(\ref{z3}) are physically-adequate and applicable on equal footing to the equations of massive particles and light in the inertial coordinates $\{t,{\bm r}\}$ whose origin is chosen at the barycenter of the solar system. Post-Newtonian corrections to the Newtonian equations are included to the computer codes to reach the computational precision being comparable with the current accuracy of astronomical observations \citep{odprogram}. The post-Newtonian corrections in the solar system ephemerides have been thoroughly studied and are well-known \citep{2011rcms.book.....K}. Hence, we do not discuss them in the present paper. We shall also neglect the motion of the solar system with respect to the cosmological reference frame defined by the congruence of the Hubble observers, and assume that the worldline of the solar system barycenter has fixed cosmological coordinates. It allows us to identify the cosmological time, $t$, with the barycentric coordinate time (TCB) used in the ephemeris astronomy \citep{2003AJ....126.2687S}. Thus, the hypersurface of simultaneity of the cosmological time $t$ coincides with that of TCB when we use the spatial coordinates ${\bm r}=a(\eta){\bm x}$.

Fitting astronomical observations to theoretical predictions utilizes the set of the Newtonian equations which leading terms are \citep{2005tvnv.conf..163S}
\ba\la{ae1}
{\bm r}''&=&-\frac{GM}{r^3}{\bm r}\;,\\\la{ae2}
\ell_{21}&=&r_{21}\;.
\ea
Here the first equation (\ref{ae1}) is used for calculating coordinates, ${\bm r}={\bm r}(t)$, of planets and spacecrafts orbiting the Sun with $M$ being the mass of the Sun. The second equation (\ref{ae2}) is the light travel time (light-cone) equation used for relating the radar distance, $\ell_{21}\equiv c(t_2-t_1)$, to the coordinate distance, $r_{21}=|{\bm r}_2-{\bm r}_1|$, between the point of emission of light with coordinates $\{t_1,{\bm r}_1\}$, and the point of reception of light at the point with coordinates $\{t_2,{\bm r}_2\}$. Light-cone equation (\ref{ae2}) can be viewed as providing a set of initial and boundary conditions for equation (\ref{ae1}). In principle, a rigorous definition of the radar distance is based on the time interval a light signal takes in order to travel from observer to a celestial target, and back. The light round-trip time is a proper time, $\t$, measured by observer equipped with atomic clock. Therefore, strictly speaking, in addition to equations (\ref{ae1}), (\ref{ae2}), we have to write down one more equation that transforms the proper time of observer, $\t$, to TCB=$t$. This equation is $\t=t+O(c^{-2})$, and is called the time ephemeris \citep{2011rcms.book.....K}. Its precise formulation and solution with the account for relativistic corrections, $\sim c^{-2}$, are well-known and has been discussed, for example, by Fukushima \citep{2010IAUS..261...89F}. Relativistic corrections in the time ephemeris equation are not important for discussing the Newtonian limit in cosmology. Hence, we omit them in the time ephemeris equation and equates $\t=t$, which is interpreted as the proper time of the Hubble observers.

The inertial coordinates $\{t,{\bm r}\}$ used in (\ref{ae1}), (\ref{ae2}) have direct physical meaning. Indeed, time $t=$TCB enters gravitational equations of motion (\ref{ae1}) and defines the ephemeris time ET which can be compared with the atomic time (AT) measured by observer on Earth. The distance $r_{21}$ is known in cosmology as the {\it proper} distance between two points lying on a hypersurface of a constant time $t$ \citep{1972gcpa.book.....W}. If the two points are infinitesimally close to each other, the proper distance $r_{21}$ coincides with the radar distance $\ell_{21}$ \citep{LL} in accordance with equation (\ref{ae2}) which is accepted as exact in the process of calculation of ephemerides. However, this equation is not true in the expanding FLRW universe. The distances $r_{21}$ and $\ell_{21}$ differ from each other, if the two points are sufficiently far away and measuring the radar distance requires integrating the light-ray equation (\ref{hus4}) along the light cone. Indeed, the Minkowskian structure of the light cone in expanding universe is preserved in the light-ray inertial coordinates $\{\lambda,{\bm r}\}$, but not in $\{t,{\bm r}\}$. Therefore, relationship (\ref{ae2}) between the radar and proper distances is only an approximation which diverges quadratically with increasing the light travel time between the two points. More specifically, equation (\ref{ae2}) should be replaced with
\be\la{ae3}
c(\lambda_2-\lambda_1)=r_{21}\;,
\ee
where
\be\la{ae4}
\lambda_1=t_1+\frac12\H t^2_1\;,\qquad\quad \lambda_2=t_2+\frac12\H t^2_2\;.
\ee
Time scale $\lambda$, is uniform for description of propagation of light while the time $t$ is a uniform time scale for celestial dynamics of massive particles. Current practice ignores the difference between the time scales, $\lambda$ and $t$. It is important to evaluate the consequences of this ignorance as it can lead to a secular divergence between predicted and observed positions and velocities of celestial bodies. It can be misinterpreted as an gravitational effect of a yet unknown force in the equations of motion of celestial bodies leading to ``anomalous orbital acceleration''  \citep{2002PhRvD..65h2004A,2010IAUS..261..189A,2008ASSL..349...75L}. We shall analyze the consequences of replacement of an approximate equation (\ref{ae2}) for exact equation (\ref{ae3}) in case of two astronomical techniques - radar/laser ranging and the Doppler tracking.

\subsection{Range Measurements}

We use equation (\ref{ae4}) and express times $t_1$, $t_2$ in terms of the coordinates ${\bm r}_1$, ${\bm r}_2$ of the points of emission and reception of the light ray by making use of equation (\ref{hus5}). Taking the Euclidean dot product of the vector equation (\ref{hus5}) with itself and accounting for the orthogonality of the impact parameter ${\bm\xi}$ and the unit vector ${\bm k}$, yield for the radar distance $\ell_{21}=c(t_2-t_1)$ the following result,
\be\la{ae5}
\ell_{21}=r_{21}+\d r_{21}\;,
\ee
where
\be\la{uit3}
\d r_{21}=-\frac{\H}{2c}\le(r^2_2-r^2_1\ri)\;.
\ee
Equation (\ref{ae5}) makes range measurement more precise, as compared with equation (\ref{ae2}), by introducing correction, $\d r_{21}$, to the proper distance, $\ell_{21}$, due to the expansion of the universe. This correction is fairly small and amounts to $\d r_{21}\simeq 10^{-7}$ cm in case of the lunar laser ranging, and $\d\ell_{21}\simeq 0.1$ cm in case of Earth-Mars radar ranging. These numbers elucidate that the expansion of the universe would not call the shots in the astronomical ephemerides if only the radar measurements were used in fitting theory to observations. However, the astronomical ephemerides also utilize another technique known as the Doppler tracking. This technique measures velocity of spacecraft which is equipped with a high-stable clock and a transponder allowing us to continuously track the phase of the Doppler signal for sufficiently long interval of time.

\subsection{Doppler tracking}

Doppler tracking is a method of measuring and control of deep space probe's velocity by making use of a radio-communication system onboard of the spacecraft via radio transmissions from Earth to spacecraft (the uplink) and returning telemetry to the ground via transmission from the spacecraft to the Earth (the downlink) \citep{odprogram,2006LRR.....9....1A}.
Analogue of the Doppler tracking in cosmological observations is the measurement of the frequency shift of a spectral line of light emitted by atoms of distant objects - galaxies and quasars - that led Edwin Hubble to the discovery of the cosmological expansion of the universe. The Hubble expansion causes shifting of the spectral lines of the light from receding cosmological objects toward the red part of electromagnetic spectrum (decrease in frequency) as compared with the laboratory spectrum. Measuring the amount of the red shift astronomers can determine a radial velocity, $v$, of the object by making use of equation for the Doppler shift \citep{lnder,1972gcpa.book.....W}. Cosmological distance, $r$, to the object relates to its radial velocity $v$ by the Hubble law which, in the first approximation with respect to the Hubble parameter, is
\be\la{ert1}
v=\H r\;.
\ee
The distance $r$ entering equation (\ref{ert1}) is the proper distance similar to definition of $r_{21}$ in the range equation (\ref{ae2}). There are other definitions of distance in cosmology based on photometric, parallactic or other kind of measurements \citep{1972gcpa.book.....W} but they coincide in the linearized Hubble law approximation (\ref{ert1}).
Spacecraft's Doppler tracking in the solar system uses a coherent measurement of phase of electromagnetic wave transmitted from Earth to the spacecraft and re-emitted back to Earth. The phase and frequency of the electromagnetic wave - both transmitted and received back - are continuously recorded by the Deep Space Network (DSN) tracking system. The Doppler tracking represents an ideal radio-communication system for performing the most precise celestial mechanical experiments with spacecrafts  \citep{2002PhRvD..65h2004A,2006LRR.....9....1A}.

We can comprehend how the cosmological expansion affects the Doppler tracking measurements by calculating the Doppler effect with the help of equation (\ref{ae3}) for the light travel time between spacecraft and observer on Earth. We denote the proper time of the atomic clock on the DSN tracking system as $\t_1$, and the proper time of clock on-board of spacecraft as $\t_2$.
The proper time $\t$ as a function of the conformal time $\eta$, conformal coordinates ${\bm x}$ and conformal velocity $\dot x^a=(c,\dot{\bm x})$, of observer/spacecraft, is defined by the integral relationship
\be\la{huw2}
\t=\int^{\eta}_{\eta_0}\sqrt{1-c^{-2}\le(\dot{\bm x}^2+h_{\a\b}\dot x^\a\dot x^\b\ri)}a(\eta')d\eta'\;,
\ee
that follows directly from the perturbed cosmological metric (\ref{i18}) and definition of the proper time, $c^2d\t^2=-g_{\a\b}\dot x^\a\dot x^\b d\eta^2$. We shall neglect the special relativistic effects and those due to the gravitational perturbation $h_{\a\b}$ which are proportional to $1/c^2$. These effects are important in high-precision space navigation but are inessential in our discussion of the influence of the cosmological expansion on the frequency of a radio wave bouncing up and down in the radio-communication link connecting Earth and spacecraft. In this approximation the proper time $\t_1$ of the terrestrial observer and that of spacecraft, $\t_2$, coincide with the TCB times of emission, $t_1$, and  reception, $t_2$, of the signal. The post-Newtonian corrections in the transformation between $\tau$ and $t$ are discussed in detail in \citep{2011rcms.book.....K}.

Let us consider, first, the up-link Doppler frequency relationship, when a radio wave is transmitted to spacecraft from the ground-based DSN antenna.
The frequency of the transmitted radio wave, $\nu_1=1/d\t_1$, where $d\t_1$ is the period of the wave. The frequency of this wave, when it reaches the spacecraft, is $\nu_2=1/d\t_2$, where $d\t_2$ is the period of the received wave. In the approximation when all terms of the order $1/c^2$ are neglected, we have $d\t_1=dt_1$ and $d\t_2=dt_2$.
The transmitted and received frequencies are linked by the light-cone equation (\ref{ae3}). Taking the total differential of this equation with respect to $dt_1$ and $dt_2$ and accounting for (\ref{ae4}), yields
\be\la{ae6}
\frac{\nu_2}{\nu_1}=\frac{dt_1}{dt_2}=\frac{1+\H t_2-{\bm k}_{21}\cdot{\bm\beta}_2}{1+\H t_1-{\bm k}_{21}\cdot{\bm\beta}_1}\;,
\ee
where the up-link unit vector ${\bm k}_{21}={\bm r}_{21}/r_{21}$ is directed from Earth to the spacecraft, ${\bm\b}_1={\bm v}_1/c$, ${\bm\b}_2={\bm v}_2/c$, ${\bm v}_1=d{\bm r}_1/dt_1$ is a velocity of the tracking station at time $t_1$, and ${\bm v}_2=d{\bm r}_2/dt_2$ is the spacecraft's velocity at time $t_2$. These velocities are modelled by solving equations of motion (\ref{ae1}) with the appropriate initial conditions \citep{odprogram}. Equation (\ref{ae6}) can be revamped by making use of the property of the unit vector: ${\bm k}_{21}\cdot{\bm k}_{21}=1$. It yields
\be\la{ae8}
\frac{\nu_2}{\nu_1}=\frac{1-{\bm k}_{21}\cdot\le({\bm\beta}_2-{\bm k}_{21}\H t_2\ri)}{1-{\bm k}_{21}\cdot\le({\bm\beta}_1-{\bm k}_{21}\H t_1\ri)}\;,
\ee
which shows that the Hubble expansion modifies velocities of observer on Earth, ${\bm\b}_1$, and spacecraft, ${\bm\b}_2$ by subtracting a small (blue shift) correction $\H t$ directed along the line-of-sight of observer toward spacecraft.

Let us assume the Earth-spacecraft distance is about the size of Pluto's orbit that is $\sim 40$ astronomical units. Then, the light travel time $t_2-t_1\simeq 2\times 10^4$ s, and the Hubble expansion yields a fractional uncertainty in the received frequency of the order $\d\nu/\nu=\H(t_2-t_1)\simeq 5\times 10^{-14}$ which is too small for detection in a single Doppler frequency measurement. However, the Doppler tracking is not a single measurement of frequency. It is actually a continuous series of coherent measurements which consists of a consecutive exchange by a radio wave bouncing between the DSN transmitter and a transponder on-board of spacecraft. The radio-tracking system observer-spacecraft makes up a radio-resonator with a high-quality factor, $Q$, and a radio wave being phase-locked inside it. Thus, the Doppler tracking technique measures an integrated change of frequency of the radio wave. This measuring process can be kept coherent for years as was demonstrated in missions Pioneer 10 and 11 \citep{2002PhRvD..65h2004A}.

The down-link of the Doppler tracking technique works as follows. Spacecraft's receiver tracks the up-link carrier frequency using a phase lock loop. The signal received on-board of spacecraft with frequency $\nu_2$ is coherently transmitted back to the ground DSN station and is received with a frequency $\nu_3$. Calculating the down-link Doppler shift with the help of the light cone equation (\ref{ae3}), we derive
\be\la{ae9}
\frac{\nu_3}{\nu_2}=\frac{1-{\bm k}_{32}\cdot\le({\bm\beta}_3-{\bm k}_{32}\H t_3\ri)}{1-{\bm k}_{32}\cdot\le({\bm\beta}_2-{\bm k}_{32}\H t_2\ri)}\;,
\ee
where the unit vector ${\bm k}_{32}={\bm r}_{32}/r_{32}$ is directed from the spacecraft to the Earth, and quantities with sub-index 3 refer to coordinates and velocity of the DSN station at the time of reception $t_3$ of the radio wave that was re-transmitted from the spacecraft toward the Earth at the time $t_2$. Technically, the procedure of re-transmission also includes multiplication of the re-transmitted frequency by a constant factor in order to prevent interference between incoming and outgoing radio signals \citep{odprogram}. We shall ignore this multiplicative factor for it changes frequency of the wave by a known number that is easily subtracted in the data processing but complicates theoretical description.

Relationship between the emitted frequency $\nu_1$ and the observed frequency $\nu_3$ is found from the condition that the phase of the radio wave in up-link is locked to that in the down-link. Hence, by applying (\ref{ae8}), (\ref{ae9}), we obtain
\be\la{ae10}
\frac{\nu_3}{\nu_1}=\frac{\nu_3}{\nu_2}\frac{\nu_2}{\nu_1}=\frac{1-{\bm k}_{32}\cdot\le({\bm\beta}_3-{\bm k}_{32}\H t_3\ri)}{1-{\bm k}_{32}\cdot\le({\bm\beta}_2-{\bm k}_{32}\H t_2\ri)}\frac{1-{\bm k}_{21}\cdot\le({\bm\beta}_2-{\bm k}_{21}\H t_2\ri)}{1-{\bm k}_{21}\cdot\le({\bm\beta}_1-{\bm k}_{21}\H t_1\ri)}\;.
\ee
Because the light travel time between Earth and spacecraft is much smaller than the orbital period of Earth around Sun, we can neglect the orbital acceleration of the Earth and use the following equalities, ${\bm k}_{32}=-{\bm k}_{21}$ and ${\bm\b}_3={\bm\b}_1$ which are valid with a very good accuracy. In fact, the acceleration terms are taken into account in real data processing but they are not important in the present paper. We substitute the above equalities to (\ref{ae10}), and expand its denominator in Taylor series. It takes (\ref{ae10}) to the following, simpler form
\be\la{ae11}
\frac{\nu_3}{\nu_1}=1-2{\bm k}_{21}\cdot\le[{\bm\b}_{2}-{\bm\b}_{1}-{\bm k}_{21}\H\le(t_3-t_1\ri)\ri]\;,
\ee
where ${\bm\b}_{2}-{\bm\b}_{1}$ is the relative radial velocity of spacecraft with respect to Earth. Introducing notation $\beta_{21}\equiv{\bm k}_{21}\cdot\le({\bm\b}_{2}-{\bm\b}_{1}\ri)$ for the radial speed of spacecraft with respect to Earth, we write equation (\ref{ae11}) as follows
\be\la{ds12}
\frac{\nu_3}{\nu_1}=1-2\le[\b_{21}-\H\le(t_3-t_1\ri)\ri]\;.
\ee
Notice that $\b_{21}>0$ when spacecraft is receding from Earth, and $\b_{21}<0$ in case when it moves toward Earth.

The up/down-link radio cycle is immediately repeated by transmitting to spacecraft at time $t_3$ of a radio wave with the same phase and frequency, $\nu_3$, as the arrived radio wave. We notice that the technical procedure used in the DSN Doppler tracking is formally different as the signal is transmitted at $t_3$ with the reference frequency $\nu_1$ (not $\nu_3$) while the difference between $\nu_3$ and $\nu_1$ is time stamped and recorded \citep{odprogram}. Since the difference between $\nu_3$ and $\nu_1$ is precisely known, we can always use in our theoretical development, frequency $\nu_3$, as that transmitted to the spacecraft at time $t_3$. This simplifies theoretical understanding of the process of accumulation of the phase of a radio wave in the two-way Doppler tracking measurement and does not change the final formula for the integrated Doppler shift.

The Doppler tracking process continues coherently following the same protocol. The wave, transmitted at time $t_3$, will be received at spacecraft at time $t_4$ with a frequency $\nu_4$, re-transmitted back to Earth with the same frequency $\nu_4$, received at DSN station at time $t_5$ with frequency $\nu_5$, transmitted back with frequency $\nu_5$ to spacecraft, and so on. Thus, the wave with a reference frequency $\nu_1$ emitted at time $t_1$ will be received at DSN station being shifted to frequency $\nu_{2n+1}$ at time $t_{2n+1}$ after $n$  up/down link bouncing cycles. Because we keep the phase of the wave locked, the relationship between the observed frequency, $\nu_{2n+1}$, and the emitted frequency, $\nu_1$, is given by the product of $n$ factors
\be\la{ae11a}
\frac{\nu_{2n+1}}{\nu_1}=\frac{\nu_{2n+1}}{\nu_{2n-1}}\frac{\nu_{2n-1}}{\nu_{2n-3}}\ldots\frac{\nu_3}{\nu_1}\;,
\ee
where the corresponding calculation of the frequency ratios is done similar to that of $\nu_3/\nu_1$. Substituting these frequency ratios to (\ref{ae11a}), summing up and neglecting the acceleration terms, yield
\be\la{ae12}
\frac{\nu_{2n+1}}{\nu_1}=1-2\sum_{i=1}^n\le[\b_{2i,2i-1}-\H\le(t_{2i+1}-t_{2i-1}\ri)\ri]\;,
\ee
where $\b_{2i,2i-1}\equiv {\bm k}_{2i,2i-1}\cdot\le({\bm\b}_{2i}-{\bm\b}_{2i-1}\ri)$ is the relative radial velocity of spacecraft with respect to Earth at time $t_{2i}$, ${\bm k}_{2i,2i-1}$ is the unit vector from the position ${\bm r}_{2i-1}$ of the DSN station taken at time $t_{2i-1}$ along the light ray toward spacecraft's position ${\bm r}_{2i}$ at time $t_{2i}$, ${\bm\b}_{2i}={\bm v}_{2i}/c$, ${\bm\b}_{2i-1}={\bm v}_{2i-1}/c$, ${\bm v}_{2i}$ is velocity of spacecraft at time $t_{2i}$ and ${\bm v}_{2i-1}$ is velocity of the DSN station at time $t_{2i-1}$.

Let us recast equation (\ref{ae12}) to the form that serves better to further discussion. We shall denote the observed frequency $\nu_{\rm obs}\equiv\nu_{2n+1}$, and equate the emitted frequency $\nu_1$ to the high-stabilized reference frequency $\nu_0$ of the master oscillator of atomic clocks, $\nu_0\equiv\nu_1$, driving the DSN radio transmitter. We shall also denote the overall time of the integrated Doppler measurement as $t=t_{2i+1}-t_1$.
The Orbit Determination Program (ODP) adopted by NASA Jet Propulsion Laboratory (JPL) models (in the approximation under consideration) the integrated Doppler shift as follows
\be\la{ae13}
\nu_{\rm mod}=\nu_0\le(1-2\sum_{i=1}^n\b_{2i,2i-1}\ri)\;.
\ee
It is more convenient for subsequent discussion to write (\ref{ae13}) in terms of an integral with respect to time by making use of the fact that the velocity of the DSN ground station and that of spacecraft do not change too much during the round trip of light from the DSN transmitter to the spacecraft and back. We can approximate the radial velocity with the increment
\be\la{az23}
\b_{2i,2i-1}=\frac{a_{\rm mod}(t'_{2i})}{c}\Delta t_{2i}\;,
\ee
where, $a_{\rm mod}=c(d\b/dt)$, is the modelled relative radial acceleration of spacecraft taken at time $t'_{2i}$ lying inside the time interval $\Delta t_{2i}=t_{2i-1}-t_{2i}$. Substituting (\ref{az23}) to (\ref{ae13}), and approximating the sum with an integral, we obtain exactly the equation used in ODP \citep{odprogram}
\be\la{ae13a}
\nu_{\rm mod}=\nu_0\le(1-\frac{2}{c}\int_0^t a_{\rm mod}(t')dt'\ri)\;,
\ee
where $t$ is the total span of observation, $t'$ is the argument of integration and, for the sake of simplicity of notations, we had started the integration at the initial epoch, $t=0$. We further notice that the accumulated phase of the radio wave (normalized to factor $2\pi$) emitted by the radio transmitter is $\phi_0=\nu_0 t$, because the frequency $\nu_0$ of the master oscillator, is constant. On the other hand, the modeled total phase of the signal, is given by the integral, $\phi_{\rm mod}=\int^t_0\nu_{\rm mod}(t')dt'$, where $\nu_{\rm mod}(t)$ is given by (\ref{ae13a}). Performing integration by parts, yields
\be\la{ae13b}
\frac{\phi_{\rm mod}-\phi_0}{\nu_0}=-\frac{2}{c}\int_0^t \le(t-t'\ri)a_{\rm mod}(t')dt'\;.
\ee

Cosmological arguments tells us that the observed Doppler frequency must be described by equation (\ref{ae12}) which includes the term depending on the Hubble expansion as a linear function of time. This prediction differs from that obtained with the JPL's ODP formula (\ref{ae13}) that was derived under assumption that the background spacetime is flat. The difference between the two equations can be written as follows,
\be\la{ae14}
\frac{\nu_{\rm obs}-\nu_{\rm mod}}{\nu_0}=2\H t\;,
\ee
where the term in the right side is obtained directly by summation of all contributions from time intervals in (\ref{ae12}): $\sum_{i=1}^n\H\le(t_{2i+1}-t_{2i-1}\ri)=\H t$. The phase difference between the observed and the ODP-modelled signals is obtained by integrating equation (\ref{ae14}) with respect to time
\be\la{ae15q}
\frac{\phi_{\rm obs}-\phi_{\rm mod}}{\nu_0}=\H t^2\;.
\ee
Both equations, (\ref{ae14}) and (\ref{ae15q}), can be formally obtained from (\ref{ae13b}) by replacing the ODP-modelled radial acceleration of spacecraft, $a_{\rm mod}$, with its cosmological counterpart,
\be\la{ae16}
a(t)=a_{\rm mod}(t)-c\H\;.
\ee
We notice that equation (\ref{ae16}) is a projection on the radius-vector connecting observer on the Earth with spacecraft. Therefore, the constant ``anomalous acceleration'', $a_{\rm anomaly}=-\H c$, of spacecraft should be interpreted as being directed toward observer but not to the Sun.

Equation (\ref{ae14}) elucidates that the contribution of the Hubble expansion to the Doppler shift of frequency grows linearly with time $t$ and, thus, can be measured after sufficiently long time interval if the phase of the radio wave used in the Doppler observation is kept locked during the time of observations (the coherent Doppler tracking). For example, during one year the fractional uncertainty between the observed and the ODP-modelled frequency as caused by the right side of equation (\ref{ae14}), amounts to $2\H\times (1\;{\rm yr})\simeq 1.5\times 10^{-10}$. Long-term stability of atomic clocks is limited by a random walk frequency noise of $\d\nu/\nu\simeq 1.0\times 10^{-16}\sqrt{t}$ where time $t$ is measured in days \citep{4319103}. For the interval of time $t =1$ year it is of $\sim 2.0\times 10^{-14}$ that is much better than the bias introduced by the Hubble expansion to the frequency drift of the coherent Doppler tracking. Analysis of the long-term stability of the coherent Doppler measurements provided by Anderson et al. \citep[section VII F]{2002PhRvD..65h2004A} yields a similar estimate for the error in measuring the left side of equation (\ref{ae14}). After comparing the estimates for the left and right sides of equation (\ref{ae14}) we conclude that the Hubble drift of the radio frequency of the observed Doppler signal is detectable with the existing telecommunication technologies.

\subsection{Does the Pioneer effect measures the Hubble constant?}
\epigraph{When you have two competing theories that make exactly the same predictions, the simpler one is the better.}{\textit{Occam's Razor Principle}}

Pioneer 10 and 11 were launched by NASA in the 1970s and radar-tracked for over 30 years. Experiment was carefully analyzed by J. Anderson with colleagues \citep{2002PhRvD..65h2004A,2007AdSpR..39..291T}. The data consistently indicated to the presence of an ``anomalous acceleration'' of spacecraft, that could be perfectly modeled by a ``phenomenological'' Doppler shift equation (see \citep[Eq. 15]{2002PhRvD..65h2004A})
\be\la{ae15}
\frac{\nu_{\rm obs}-\nu_{\rm mod}}{\nu_0}=\frac{2a_P t} {c}\;,
\ee
with $a_P=(8.74\pm 1.33)\times 10^{-8}$ cm/s$^2$. It was noticed that the value of $a_P$ is very close to $\H c$ pointing out to possible explanation of the ``anomalous acceleration'' by the Hubble expansion. However, the sign of the effect (blue shift) was opposite to that having been naively expected from cosmology (red shift), and it was the main reason why all previous theoretical attempts to explain the blue shift of the Pioneer effect by cosmological expansion, have failed  \citep{2005AmJPh..73.1033T,2010RvMP...82..169C}. Meanwhile, experimentalists were trying to explore more deeply the hypothesis that the anisotropic heat emitted by the spacecraft's generators is the primary cause of the effect \citep{2011PhRvL.107h1103T,2011AnP...523..439R} in spite of the original paper by Anderson et al. \citep{2002PhRvD..65h2004A} had presented convincing arguments that the heat (and the other on-board generated systematics) cannot be responsible for the overall effect but only for its small fraction ($\sim 15\%$ of the effect). Nevertheless, the efforts to explore the efficiency of the radiative force-recoil model continue. I am aware (private communication) of at least three works that are currently in progress, not as of this date announced or published, that in effect claim that the Pioneer anomaly can be explained in terms of heat radiation by the spacecraft. Definitely, the heat of the on-board generators should produce some effect on the orbital acceleration of the spacecraft. However, the present paper points out that there is a strong theoretical argument in favor that the prime cause of the Pioneer anomaly is deeply rooted in the fundamental gravitational physics of spacetime rather than in the thermal process. 

The theory of various time scales worked out in the present paper, is able to explain the Pioneer effect as caused by a secular drift of the frequency of a radio wave that is phase-locked in the DSN radio communication system observer-spacecraft. The drift of the radio wave frequency is caused by the cosmological expansion that affects the rate of electromagnetic time scale, $\lambda$, as compared with that of the atomic and ephemeris time scales, $t$, (see equation (\ref{hus9})). It explains both the magnitude of the frequency drift and the blue-shift sign of its time derivative. The effect is absent in Minkowski spacetime where all time scales (atomic, ephemeris, and electromagnetic) measured by a set of fundamental observers, have the same rate.

It is worthwhile to notice that \citep{2002PhRvD..65h2004A} attempted to fit the Pioneer effect by introducing a quadratic time divergence of atomic clocks from the ephemeris time of gravitational equations of motion. This did not work out and this negative result is in a full accordance with our prediction that the rate of atomic and ephemeris time scales have no divergence in expanding universe.  On the other hand, Anderson et al. \citep{2002PhRvD..65h2004A} were successful to fit the Pioneer effect with the {\it phenomenological} quadratic-in-time model that was applied only in equations of light propagation. However, neither at that time nor later the {\it quadratic-in-time} model did not find theoretical justification and remained purely phenomenological toy-model \citep{2008ASSL..349...75L}. Our theory provides a solid theoretical framework supporting the quadratic-in-time model originating from the non-uniformity of the electromagnetic time scale, $\lambda=t+1/2\H t^2$, realized by a photon bouncing between two 'radio mirrors' (observer-spacecraft) in an expanding universe. This non-uniformity of electromagnetic time scale leads unequivocally to the Doppler shift equation (\ref{ae14}) and provides theoretical explanation to the experimentally-established equation (\ref{ae15}).

Equation (\ref{ae14}) can be also reformulated in the form of (\ref{ae16}), and interpreted in terms of the constant ``anomalous acceleration'', $a_{\rm anomaly}=-\H c$, of spacecraft directed toward observer. If spacecraft is moving away from observer the ``anomalous acceleration'' leads to a tiny blue shift of the Doppler frequency of the spacecraft's signal on the top of a large red shift caused by the outgoing motion of spacecraft. We also notice that equation (\ref{ae16}) is a projection on the line connecting observer with spacecraft. Therefore, the ``anomalous acceleration'' is directed toward observer on the Earth but not to the Sun. It agrees with the observed direction of the ``anomalous acceleration'' \citep{2002PhRvD..65h2004A}. Equation (\ref{ae16}) also tells us that the ``anomalous acceleration'' is constant everywhere within the solar system and far outside of its boundaries. This is because the ``anomalous acceleration'' is not a physical force but the effect of the cosmic expansion which is homogeneous and isotropic everywhere in the FLRW universe.

The ``anomalous acceleration'' is the most common {\it phenomenological} interpretation of the Pioneer effect adopted in most literature \cite{2008ASSL..349...75L,2010RvMP...82..169C}. We emphasize that our theory associates the physical origin of the Pioneer effect neither with the ``anomalous acceleration'' nor with an ``anomalous gravitational force'' but with the different rates of the time scales caused by the cosmological expansion. The search for the ``anomalous gravitational force'' was a matter of many theoretical speculations aimed to explain the Pioneer effect with a ``new physics beyond Einstein'' \citep{2008ASSL..349...75L}. Our standpoint is that the Pioneer effect can be explained within the framework of general relativity as a natural consequence of the (locally-observed) Hubble expansion without resorting to alternative theory of gravities or other exotic phenomena.

Comparing (\ref{ae15}) with (\ref{ae14}) yields a theoretical prediction for $a_P=\H c$. Taking the WMAP value for $\H$ \citep{0067-0049-192-2-14} leads to $a^{\sst WMAP}_P=(6.9\pm 0.3)\times 10^{-8}$ cm/s$^{2}$.  On the other hand, substituting the numerical value for $a_P=(8.74\pm 1.33)\times 10^{-8}$ cm/s$^2$, measured in Pioneer experiment \citep{2002PhRvD..65h2004A}, yields a corresponding value of the Hubble constant $\H=90.0\pm 13.7$ km/s/Mpc that is larger than the value of $\H=71.0$ km/s/Mpc derived from WMAP data.  More recently obtained value of $a_P=(8.17\pm 0.02)\times 10^{-8}$ sm/s$^2$ \citep{2011PhRvL.107h1103T} yields, $\H =84.2\pm 0.2$ km/s/Mpc, that still noticeably exceeds the WMAP value. The difference is about $20\%$ and may be explained, indeed, by accounting for the residual thermal emission of spacecraft's generator.

Cassini spacecraft was also equipped with a coherent Doppler tracking system and it might be tempting to use the Cassini telemetry to measure the ``anomalous acceleration effect'' which is predicted by the same theoretical equation (\ref{ae16}) and, if WMAP value for $\H$ is implemented, has the magnitude of $\sim 7$ cm/s$^{2}$. Unfortunately, there are large thermal and outgassing effects on Cassini that would make it difficult or impossible to say anything about the ``Pioneer anomaly'' from Cassini data, during its cruise phase between Earth and Saturn \citep{2010PEPI..178..176A}. Due to the presence of the Cassini on-board generated systematics, the recent study \citep{2012arXiv1201.5041H} of radioscience simulations in general relativity and in alternative theories of gravity is consistent with a non-detection of the ``anomalous acceleration'' effect.

\section{The Concluding Remarks}\la{cremar}
\subsection{Cosmological Effects in Local Inertial Coordinates}
\epigraph{There are more things in heaven and earth, Horatio, than are dreamt of in your philosophy.}{Shakespeare.\textit{ Act I, scene {\rm v}, Hamlet}}

The reader may be wondering how it is possible that the Hubble expansion can be measured locally. It seems to be in contradiction to a common wisdom which states that any curved manifold admits a construction of local inertial coordinates along any time-like worldline such that all effects produced by the manifold's curvature vanish at the origin of the coordinate chart. This is because the Christoffel symbols are nil at the origin of the local inertial coordinates \citep{mtw}.

The present paper does not argue against this well-established mathematical fact. The local inertial coordinates in cosmological spacetime can be built in accordance with the mathematical principles of the Riemannian geometry \citep{2006PhRvD..74f4019C,2005ESASP.576..305K,hongya:1920,hongya:1924}. The question is not about mathematics but about how the coordinate description of motion of test particles and light relates to physical observables -- the proper time and frequency of light -- measured by an observer at the origin of the local inertial coordinates. Physical treatment of this problem in the linearized Hubble approximation is given below. Many technical complications will be passed over. More specifically, we shall drop out terms of the order of $\H^2$ from equations. We shall also neglect the effects of the gravitational field of the localized astronomical system which are incorporated to the metric tensor perturbations $l_{\a\b}$. They can be easily added up by making use of the principle of linear superposition as confirmed by calculations given in sections \ref{pnfec}, \ref{nlfe} of the present paper.

We start from the FLRW metric (\ref{i3}) and make transformation (\ref{hus1}) of spatial coordinates, $r^i=R(t)x^i$. We obtain the FLRW metric in the new coordinates,
\be\la{fgh3}
ds^2=-c^2dt^2-2\H r^i dtdr^i+\d_{ij}dr^i dr^j\;,
\ee
where all terms of the order of $\H^2$ have been omitted. We remind that the cosmological time $t$ is the proper time of the Hubble observers having fixed spatial coordinates $x^i={\rm const.}$. Clocks of the Hubble observers are synchronized with the help of Einstein synchronization procedure based on exchange of electromagnetic signals \citep{LL,1972gcpa.book.....W}, and the rate of the clocks is uniform.

Strictly speaking, coordinates $\{t,r^i\}$ are not locally inertial because not all Christoffel symbols in these coordinates vanish. In particular, the only non-vanishing Christoffel symbol is $\bar\G^0{}_{ij}=(\H/c)\d_{ij}$. This Christoffel symbol causes a non-inertial force and is responsible for the accelerated motion of photons in coordinates $\{t,r^i\}$ as shown in equation of motion (\ref{hus4}). On the other hand, the non-vanishing Christoffel symbol affects equations of motion of slowly moving test particles only in the post-Newtonian terms of the order of $\H/c^2$ which are negligibly small in the solar system. This is the reason why the coordinates $\{t,r^i\}$ can be treated as inertial in the Newtonian limit of the slowly moving particles. It corresponds to the discussion given in section \ref{hjqcv}.

Metric (\ref{fgh3}) can be further simplified by making use of time transformation,
\be\la{fghas}
\lambda=t+\frac1c\bar\G^0{}_{ij}r^ir^j+\ldots\;,
\ee
where  $\lambda$ is a new coordinate time, and the dots denote residual terms of the higher order in the spatial coordinates and the Hubble parameter. After substituting the above-given value for $\bar\G^0{}_{ij}=(\H/c)\d_{ij}$, the explicit form of the coordinate transformation becomes
\be\la{fgh4}
\lambda=t+\frac{\H}{2c^2}r^2\;,
\ee
where the residual terms have been dropped out. The reader should notice that equation (\ref{fgh4}) apparently indicates that the spatial hypersurfaces of constant time, $\lambda$, do not coincide with the hypersurfaces of simultaneity of the proper time $t$ of the Hubble observers.

The metric (\ref{fgh3}) has in the new coordinates $\{\lambda,r^i\}$, apparently Minkowskian form
\be\la{fgh5}
ds^2=-c^2d\lambda^2+\d_{ij}dr^i dr^j\;.
\ee
It reveals that coordinates $\{\lambda,r^i\}$ are locally inertial coordinates introduced on the cosmological FLRW spacetime manifold in the neighborhood of the worldline of a Hubble observer having fixed spatial coordinates $r^i=0$. All Christoffel symbols disappear at the origin of the local inertial coordinates $\{\lambda,r^i\}$ that can make an impression that all cosmological effects which might be measured are shifted to the residual terms of the order of $\H^2$. This conclusion is invalid.

The fact is that the coordinate time $\lambda$ is not the proper time $t$ of the Hubble observer outside of the observer's worldline, and is not directly measurable. Therefore, the coordinate description of motion of test particles with respect to the coordinate chart, $\{\lambda,r^i\}$, is not given in terms of physically-measurable time, $t$, and should not be interpreted like the motion of particles in Minkowskian spacetime. To make this important point more clear, let us consider a test particle moving freely from the origin of the local inertial coordinates. Equation of motion of the particle is geodesic, which is given in the metric (\ref{fgh5}) by the first Newton's law
\be\la{fgh6}
\frac{d^2r^i}{d\lambda^2}=0\;.
\ee
This equation can be integrated easily
\be\la{fgh7}
r^i=v^i\lambda\;,
\ee
where $v^i=dr^i/d\lambda$ is a constant velocity of the particle. We have to replace the unobservable coordinate time, $\lambda$, with the directly-measurable proper time $t$ of the observer which is located at the origin of the coordinates. Relationship between $\lambda$ and $t$ is obtained from the coordinate transformation (\ref{fgh4}) which must be performed at the current position of the particle given by equation (\ref{fgh7}),
\be\la{fghokn}
\lambda=t+\frac12\H\frac{v^2\lambda^2}{c^2}\;.
\ee
Solving by iterations, we get
\be\la{fgh8}
\lambda=t+\frac12\H\frac{v^2t^2}{c^2}+\ldots\;,
\ee
where the dots denote residual terms of the higher order in $t$. Equation (\ref{fgh8}) demonstrates that the coordinate time, $\lambda$, diverges quadratically from the proper time, $t$, of the observer as the particle moves away from the origin of the local coordinates. The divergence between the two time scales, $\lambda$ and $t$, is a reflection of the symmetry of the underlying cosmological manifold pointing out that the {\it coordinate} simultaneity in the local inertial coordinates does not correspond to the {\it physical} simultaneity of clocks of the Hubble observers in the expanding universe.

The divergence between $\lambda$ and $t$ is negligibly-small for slowly-moving particles due to the presence of the post-Newtonian parameter, $(v/c)^2\ll 1$, in combination with the Hubble parameter $\H$. However, if the particle is a photon (radio wave), its velocity $v=c$, and equation (\ref{fgh8}) becomes
\be\la{fgh9}
\lambda=t+\frac12 \H t^2\;.
\ee
This equation is identical with (\ref{hus9}) and elucidates our derivation of the propagation of light given in section \ref{emli}. We conclude that the Hubble parameter enters equations of motion of test particles in the local inertial coordinates in cosmology. Time coordinate, $\lambda$, is a convenient mathematical parameter along the path of test particles but it must be expressed in terms of the observable proper time $t$ of the Hubble observer in order to make physical sense of the parametrization.

\subsection{Experimental prospects to measure the Hubble expansion in the solar system}\la{expHu}

Pioneer effect caused a keen interest in scientific community of experimental and gravitational physicists. There were several proposals submitted to NASA and ESA \citep{2009ExA....23..529C,2007AdSpR..39..291T} describing dedicated space missions to test the ``anomalous force'' and ``celestial anomalies'' in the solar system. All proposed missions requested the launching spacecraft to deep space beyond the orbit of Jupiter.

According to our theory, the testing of the ``anomalous force'' has nothing to do with the violation of general relativity. The effect of the ``anomalous force'' detected in Pioneer 10 and 11 missions, is associated with the non-uniformity of electromagnetic time scale (Marzke-Wheeler clock \citep{marweel}) as compared with atomic time scale (atomic clock) and the ephemeris time scale (gravitational clock). Therefore, the effect can be tested in any space mission having highly stabilized atomic clock which long-term stability is measured with a coherent Doppler tracking system. The most restrictive requirements to the mission are:
\begin{enumerate}
\item the suppression and/or precise calibration of the on-board systematics like heat, gas leakage, etc.,
\item the reduction of the impact (or precise measurement with on-board detectors) of the non-gravitational forces caused by the solar wind, light pressure, dust, etc.,
\item deployment of ranging measurement technique that ensures an independent determination of the space probe position and velocity.
\end{enumerate}
PHARAO project \citep{2007SPIE.6673E...6L} may be an excellent candidate for measuring the effect of the local cosmological expansion which has a well-predicted magnitude, sign and direction in accordance with our equation (\ref{ae14}). Another possibility is to deploy a radio transmitter in a package with laser retro-reflector on the visible face of the Moon in order to conduct the coherent Doppler and laser ranging measurements to test (\ref{ae14}). It can be also achieved with the high-precision Doppler and satellite laser ranging of a telecommunication spacecraft launched at a geostationary orbit. Constellation of GPS satellites may be useful as well but this should be explored more deeply as the GPS satellite orbit degrades as time goes on.

It may happen that measuring the local value of the Hubble constant can be achieved with the optical flywheel oscillator \citep{Tobar5168185}. The oscillator represents an optical resonator whose current instability is $\sim 3.5\times 10^{-14}t^{-1/2}$. The frequency of this type of oscillator, as compared with that of atomic clocks, is expected to drift due to the Hubble expansion as $\H t$ in accordance with (\ref{ae14}). In principle it can be measured already after 1000 seconds. However, the theory of the effect of the Hubble local expansion in optical oscillators should be worked out in addition to the calculations of the present paper. Moreover, frequency instability of optical oscillator is caused by many other factors which should be carefully studied and subtracted.

\acknowledgments
\epigraph{My friends were poor but honest.}{Shakespeare.\textit{Act I, scene {\rm iii}, Helena}}

{This paper was made possible through the support of a grant from the John Templeton Foundation. The opinions expressed in this publication are those of the authors and do not necessarily reflect the views of the John Templeton Foundation. The funds from John Templeton Foundation were awarded in a grant to the University of Chicago which also managed the program in conjunction with National Astronomical Observatories, Chinese Academy of Sciences.

This research was partially supported by the Chinese Academy of Sciences Visiting Professorship for International Senior Scientists.

The author is thankful to the International Space Science Institute (Bern, Switzerland) for hospitality and travel support to present the results of the present paper at the 3-d Lunar Laser ranging workshop (March 22-23, 2012), and to the participants of the workshop for discussion. 

Ignazio Ciufolini, Adam Helfer, Haojing Yan and Vladimir Ilchenko made a number of valuable comments which helped to improve the paper. Fruitful scientific discussion with Alexander Petrov, Bahram Mashhoon, John D. Anderson, Allen J. Anderson, and Haojing Yan is deeply appreciated. 

The author is grateful to anonymous referee for stimulating report which helped to clarify the presentation of the physical ideas and theoretical results in the manuscript.
}

\bibliographystyle{plain}
\bibliography{NF-cosmology}

\end{document}